\documentclass[10pt,twocolumn,a4paper]{IEEEtran}

\usepackage[active]{srcltx} 
\usepackage[cmex10]{amsmath}
\usepackage{amsfonts}
\usepackage[final]{graphics}
\usepackage[final]{graphicx}
\usepackage{amsbsy}
\usepackage{amsbsy}
\usepackage{amssymb}
\usepackage{url}
\usepackage{enumerate}
\usepackage{cite}
\usepackage{psfrag}
\usepackage{color}
\usepackage{euscript}
\usepackage{float}
\usepackage[utf8]{inputenc}
\usepackage{algorithmic}
\usepackage[ruled,vlined]{algorithm2e}
\usepackage{array,amsthm}
\usepackage{gensymb}       
\usepackage[table]{xcolor}
\usepackage{multirow}
\usepackage{booktabs}  
\usepackage{gensymb}
\newcommand{\ignore}[1]{}  
\usepackage{mathtools}
\usepackage{cite}
\usepackage{textcomp}
\usepackage{siunitx}
\usepackage{xcolor}
\usepackage[]{mdframed}
\usepackage{rotating}
\usepackage[export]{adjustbox}
\def\BibTeX{{\rm B\kern-.05em{\sc m\kern-.025em b}\kern-.08em
    T\kern-.1667em\lower.7ex\hbox{E}\kern-.125emX}}

\newtheorem{definition}{Definition}

\newcommand{\bm}[1]{{\mathbf{#1}}}     
\newcommand{\Es}{{\mathbb{E}}}         
\newcommand{\eqdef}{\triangleq}
\newcommand{\card}{\text{card}}
\newcommand{\herm}{\text{H}}
\newcommand{\trasp}{\text{T}}

\newcommand{\Cset}{\mathbb{C}}
\newcommand{\Rset}{\mathbb{R}}

\newcommand{\Zset}{\mathbb{Z}}
\newcommand{\capa}{\EuScript{R}}
\newcommand{\capaa}{\EuScript{C}}
\newcommand{\bC}{\bm C}

\newcommand{\bc}{\bm c}

\newcommand{\bv}{\bm v}
\newcommand{\energy}{\EuScript{E}}
\newcommand{\fair}{\EuScript{F}}

\def\be#1\ee{\begin{equation}#1\end{equation}}
\def\barr#1\earr{\begin{align}#1\end{align}}

\usepackage{tikz}

\newcommand\copyrighttext{%
  \footnotesize \textcopyright \the\year{} IEEE. Personal use of this material is permitted. Permission from IEEE must be obtained for all other uses, including reprinting/republishing this material for advertising or promotional purposes, collecting new collected works for resale or redistribution to servers or lists, or reuse of any copyrighted component of this work in other works.}

\newcommand\copyrightnotice{%
\begin{tikzpicture}[remember picture,overlay]
\node[anchor=south,yshift=10pt] at (current page.south) {%
\begin{minipage}{\textwidth}
\center \copyrighttext
\end{minipage}};
\end{tikzpicture}%
}

\newcommand\acceptedtext{%
  \footnotesize This article has been accepted for publication in a future issue of this journal,
  but has not been fully edited. Content may change prior to final publication. \\
  Citation information: DOI 10.1109/TVT.2024.3432170, IEEE Transactions on Vehicular Technology.}
  
\newcommand\acceptednotice{%
\begin{tikzpicture}[remember picture,overlay]
\node[anchor=north,yshift=-6pt] at (current page.north) {%
\begin{minipage}{\textwidth}
\center \acceptedtext
\end{minipage}};
\end{tikzpicture}%
}

\begin{document}

\title{A Hybrid NOMA-OMA Scheme for Inter-plane Intersatellite Communications
in Massive \\ LEO Constellations
}
\author{Donatella~Darsena,~\IEEEmembership{Senior Member,~IEEE},
Giacinto~Gelli,~\IEEEmembership{Senior Member,~IEEE},
Ivan~Iudice,
and
Francesco~Verde,~\IEEEmembership{Senior Member,~IEEE}
\thanks{
Manuscript received December 5, 2023; 
revised May 22, 2024;
accepted July 16, 2024.
The associate editor coordinating the review of this paper and
approving  it for publication  was Prof.~Zhu Han.
(\em Corresponding author: Francesco Verde)
}
\thanks{D.~Darsena, G.~Gelli, and F.~Verde are with the Department of Electrical Engineering and Information Technology,  University Federico II, Naples I-80125,
Italy [e-mail: (darsena, gelli, f.verde)@unina.it].
I.~Iudice is with the Reliability \& Security Department, 
Italian Aerospace Research Centre (CIRA), Capua I-81043, Italy
(e-mail: i.iudice@cira.it).
}
\thanks{
D.~Darsena, G.~Gelli, and F.~Verde are also with
National Inter-University Consortium for Telecommunications (CNIT).}
\thanks{This work was partially supported by the European Union under the Italian National Recovery and Resilience Plan (NRRP) of NextGenerationEU, partnership on ``Telecommunications of the Future" (PE00000001 - program ``RESTART").}
\thanks{This research activity falls within the field of interest of the IEEE AESS
technical panel on Glue Technologies for Space Systems.}
}
\markboth{IEEE Transactions on Vehicular Technology, Vol.~xx,
No.~yy,~zz~2024}{Darsena \MakeLowercase{\textit{et al.}}:
A Hybrid NOMA-OMA Scheme for Inter-plane Intersatellite Communications
in Massive LEO Constellations}

\IEEEpubid{0000--0000/00\$00.00~\copyright~2024 IEEE}

\maketitle
\copyrightnotice
\acceptednotice

\begin{abstract}
Communication between satellites in low-Earth orbit (LEO) 
constellations takes place through inter-satellite links (ISLs).
Unlike intra-plane ISLs, which interconnect satellites belonging 
to the same orbital plane with fixed 
relative distance, inter-plane ISLs experience significant 
Doppler frequency shifts, since satellites 
belonging to different orbital planes exhibit 
time-varying relative distance (required, e.g., to minimize 
the risk of physical collisions between satellites). 
In this paper, we consider the problem of connecting 
multiple satellites, belonging to a massive LEO 
constellation, to a receiving satellite, 
referred to as the sink.
Specifically, we consider a hybrid 
multiple access scheme, which employs 
a combination of non-orthogonal multiple access (NOMA), 
where radio-frequency ISLs share the same time-frequency resource blocks,
and orthogonal multiple access (OMA), where ISLs
employ orthogonal resource blocks.
The set of satellites 
transmitting towards the sink
is divided into groups, where NOMA 
is employed within each 
group, whereas OMA is used to separate 
different groups.
Such a scheme subsumes as special cases both pure-OMA and pure-NOMA.
Our study highlights that similar Doppler frequency shifts 
have a significant impact on the individual rates of the satellites in a pure-NOMA scheme, 
thus reducing the network fairness of this technique.
Motivated by such a fact, we develop design strategies of the proposed
hybrid NOMA-OMA scheme, which exploit
inter-plane Doppler frequency diversity 
to enhance fairness among
the satellites, while ensuring a significantly higher sum-rate capacity
compared to the pure-OMA technique.
Numerical results corroborate our theoretical analysis, by demonstrating
both the fairness enhancement of the proposed techniques  over the 
pure-NOMA scheme, as well as their capacity 
improvement over the pure-OMA one.

\end{abstract}


\section{Introduction}
\label{sec:intro}


\IEEEPARstart{T}{here} is an upsurge of interest,
both in the industrial
and academic community, towards
\textit{low-Earth orbit}
(LEO) satellite constellations,
whose satellites are deployed
between $400$ and $2000$ km over the Earth surface.
Compared to higher orbits,
LEO constellations
offer better coverage and reduced propagation delays.
Such advantages have spurred standardization
activities, e.g.,  within 3GPP,
to integrate LEO satellite networks
into 5G and beyond-5G networks
\cite{LeyvaMayorga2020,Darwish2022}.

In addition to national/international public institutions,
an increasing number of private companies 
(e.g., SpaceX, OneWeb, Amazon, and Google)
have planned and started to deploy
LEO constellations composed by
thousands of satellites -- so called \textit{megaconstellations} -- 
with the aim of providing global services, such as
broadband Internet access and mobile telephony,
as well as to support Internet-of-Remote-Things (IoRT) applications \cite{Chu2021}.
In such megaconstellations, satellites
are typically organized
in groups, each one following the
same orbital trajectory, which lies in a plane
called \textit{orbital plane} (OP).
The angle between the OP and the equatorial plane is called
\textit{inclination}, which ranges 
from $0\degree$ (equatorial orbits)
to nearly $90\degree$ (polar orbits).
For instance, the Starlink megaconstellation envisioned
by SpaceX is expected to
be comprised by
approximately $12,000$ LEO and
very LEO (i.e., under $400$ km) satellites,
to be deployed in two phases.
In phase I, SpaceX is deploying
a LEO constellation of $1584$ satellites,
arranged in $22$ orbital planes (OPs),
with $72$ satellites per OP,
at an altitude of $550$ km and an inclination
of $53\degree$.

\IEEEpubidadjcol


In next-generation LEO satellite networks,
in order to satisfy 
more challenging performance requirements,
in terms of connectivity, coverage, capacity, and latency,
it is commonly recognized \cite{Chaudhry2021}
that point-to-point \textit{intersatellite links} (ISLs)
will play an important role.
ISLs can be classified 
\cite{Chaudhry2021,Chaudhry2021a} in two main categories:
(1) \textit{intra-plane}, which are established 
between two satellites belonging 
to the same OP;
(2) \textit{inter-plane}, which are established between two 
satellites belonging to different OPs. 
Moreover, inter-plane ISLs
can be of three types \cite{Chaudhry2021a}: between adjacent OPs, between nearby OPs, 
between crossing OPs.

Both free-space optical (FSO) 
and radio-frequency (RF) 
communication technologies 
can be considered 
for ISL implementation
\cite{Al-Hraishawi.2023, Leyva-Mayorga.2020, Nie.2021}.
The FSO solution employs very narrow beams to counteract 
the detrimental effects of path loss at high 
carrier frequencies, while 
assuring in principle high data-rates, 
large transmission ranges, and low sensitivity 
to interference. 
While optical intra-plane ISLs are easy
to be established and managed, due to the
slow relative motion between satellites,
inter-plane optical ISLs demand significant challenges to acquisition, 
tracking and pointing (ATP) on-board systems, 
due to the ultra-narrow beamwidth of FSO links and the high
relative speeds of the satellites \cite{Chaudhry2021a}.
In contrast, in inter-plane ISLs, 
the lower operation frequencies of radio ISL 
enable easier beam alignment and 
allow one to connect  with a higher 
number of neighbor satellites, 
along with an easier integration into terrestrial 
RF-based networks.
Among the disadvantages, radio ISLs are
more susceptible to interference 
and provide lower data-rate 
compared to optical links, besides
requiring a licensed bandwidth to operate.
Moreover, inter-plane radio 
ISLs, especially those between crossing OPs, 
exhibit high Doppler frequency shift values, 
which become one of the main performance 
limiting factor \cite{LeyvaMayorga2021}.
It is worthwhile to note that the RF technology may be a 
reliable fallback solution when FSO links are unavailable, 
thereby making hybrid RF/FSO systems good candidates for
inter-satellite networking  \cite{Leyva-Mayorga.2020}.

For the above mentioned reasons, 
we will consider in the following
innovative designs of inter-plane radio ISLs
in a LEO network scenario.
Indeed, traditional ISL designs 
for geostationary orbit
(GEO) systems must be rethought, 
to be adapted 
to the highly dynamic features
of LEO satellite networks, which are
characterized by  high mobility, coupled with 
relatively low-power transmission 
capabilities.
%
%
%
%
%
Specifically, to avoid interference between ISLs 
caused by the sharing of RF  resources
by multiple communicating satellites \cite{LeyvaMayorga2021},
state-of-the-art satellite communications
mainly adopt \textit{orthogonal multiple access} (OMA) schemes,
such as frequency-division multiple access (FDMA),
time-division multiple access (TDMA),
code-division multiple access (CDMA),
orthogonal frequency-division multiple access (OFDMA),
or combinations thereof.
In recent years,
\textit{non-orthogonal multiple access} (NOMA)
has been recognized \cite{Islam2017} as a promising
approach for terrestrial networks,
due to its superior spectral
efficiency with respect to OMA schemes.
Indeed, in NOMA,
multiple users are allowed to
share the same time-frequency resource (or pool of resources),
through power-domain or code-domain multiplexing.
The superimposed signals can be separated at the receiver
by using successive interference cancellation (SIC).
At the cost of an increased receiver complexity, NOMA techniques
offer not only improved spectral efficiency,
but also fairness in some cases \cite{Shin2017}.
Spectrum efficient and
secure transmission schemes for satellite communication systems
have been proposed and analyzed in \cite{Yin_2019,Yin_2023}, where 
terrestrial nodes access a satellite downlink network
with frequency-division NOMA.

This paper focuses on design 
of hybrid NOMA-OMA 
multiple access strategies for
inter-plane radio ISLs to be employed in LEO megaconstellations.
Instead of marking ISLs with high Doppler 
shifts as non-feasible \cite{LeyvaMayorga2021,Pi2022a},
our approach opportunistically leverages 
the different Doppler profiles 
of the satellites (so called \textit{Doppler diversity})
to achieve a performance gain, as explained in the following.

\begin{table*}
\caption{Main parameters and symbols (the time variable $t$ is omitted).}
\label{tab:main}
\centering{}%
\begin{tabular}{clcl}
\hline
\noalign{\vskip\doublerulesep}
\textbf{Symbols} & \textbf{Meaning} & \textbf{Symbols} & \textbf{Meaning} \tabularnewline[\doublerulesep]
\hline
$K$  &  number of satellites & $\mathcal{L}$  &  set of indices of feasible links 
\tabularnewline
\hline
$P$  &  number of OPs & $\mathbb{P}({\mathcal{L}})$ & partition of $\mathcal{L}$ (collection of satellite groups) 
\tabularnewline
\hline
$N$  &  number of satellites for each OP &$G$ & number of satellite groups
\tabularnewline
\hline
$h$  &  satellite's altitude &$\rho_{k}$ & fraction of DoF allocated to $\mathcal{L}_k$
\tabularnewline
\hline
$R$  &  Earth's radius &$A_{\ell_k,q}$ & amplitude of the link between 
sink and $(\ell_k,q)$-th satellite
\tabularnewline
\hline
$\theta_{p,n}$  & colatitude of the $(p,n)$-th satellite &$\tau_{\ell_k,q}$ & delay of the link 
between  sink and  $(\ell_k,q)$-th satellite
\tabularnewline
\hline
$\phi_{p,n}$  & longitude of the $(p,n)$-th satellite &$\nu_{\ell_k,q}$ & normalized Doppler of the link between sink and $(\ell_k,q)$-th satellite
\tabularnewline
\hline
$d_{p,n}$  & distance between sink and $(p,n)$-th satellite  &$T$ & symbol period
\tabularnewline
\hline
$L$  &  number of feasible links &$T_c$ & sampling period
\tabularnewline
\hline
$S$  & oversampling factor &$N_0$ & observation interval length (in symbol periods)
\tabularnewline
\hline
$\text{P}_{p,n}^{\text{rx}}(t)$  &  power transmitted by the $(p,n)$-th satellite
& $\text{G}_{p,n}^{\text{tx}}$ & antenna gain of the $(p,n)$-th transmitting
satellite
\tabularnewline
\hline
$\mathbf{r}_{p,n}(t)$ &  ECEF position of the $(p,n)$-th satellite
& $\text{G}_{1,1}^{\text{rx}}$ & antenna gain of the sink
satellite
\tabularnewline
\hline
\end{tabular}
\end{table*}

\subsection{Related work and open research}
\label{sec:related}

Some works have considered NOMA for satellite networks
(see \cite{Yan2019} and references therein), but are mainly
limited  to air-to-ground (downlink)
or ground-to-air (uplink) communications.
A power-domain NOMA scheme
for the downlink of an integrated
satellite-terrestrial
network was considered in \cite{Chu2021},
targeted at Internet-of-Things (IoT) applications.
With particular reference to LEO constellations, 
the performance analysis of downlink NOMA
is carried out in
\cite{Gao2020}, in terms of ergodic capacity,
outage probability, and mutual information,
by considering the case of
two ground users in one spot beam.
In \cite{Gao2021}, a sum-rate capacity maximization
problem was considered
for a LEO constellation employing massive
MIMO in conjunction with NOMA.
In \cite{Ge2021}, a NOMA scheme has been introduced
in uplink,  in the context of a multi-layer satellite networks, composed by
GEO and LEO satellites.
The works \cite{Chu2021,Gao2020,Gao2021,Ge2021}
do not consider intersatellite (IS) communication at all.

The problem of IS communication has been studied
in \cite{Tachikawa2021}, where communications between two satellite
groups are established with the help of
a relay satellite and applying NOMA schemes for both downlink and uplink.
NOMA schemes for a dual-layer LEO megaconstellation
are studied in \cite{Hu2023}, 
aimed at allowing the coexistence of services with different requirements.
However, in \cite{Tachikawa2021,Hu2023}, Doppler frequency shifts are regarded as
a cause of performance degradation rather than as a source of diversity.

The design of ISLs for dense LEO constellations has been considered
in \cite{LeyvaMayorga2021}, where the establishment of
inter-plane ISLs is casted as a weighted dynamic matching problem
and solved by performing sum-rate capacity maximization.
However, this research considers an OMA scheme and,
moreover, marks as \textit{non-feasible} those links characterized
by high Doppler frequency shift values.
Another work in this area is \cite{Pi2022a},
which deals with the problem of designing inter-plane ISLs, based on
reinforcement learning, by discarding links
affected by non-negligible Doppler effects.

\subsection{Contributions}
\label{sec:contribution}

Our contributions can be summarized as follows.

\begin{enumerate}[1)]

\item
We provide an accurate characterization of radio ISLs 
that allow multiple satellites of a LEO Walker Delta constellation to effectively communicate with
a receiving satellite, referred to as the \textit{sink}. In particular, we unveil 
that the satellites with feasible ISLs towards the sink belong not only 
to the same OP of the sink (intra-plane), but also to adjacent 
and crossing OPs (inter-plane), exhibiting thus significant Doppler frequency shifts.  

\item
We show that the Doppler frequency shifts characterizing 
inter-plane radio ISLs have a significant 
impact on the individual rates of the satellites 
that access the channel in a non-orthogonal manner 
with minimum mean-squared error (MMSE)-plus-SIC 
receiver at the sink, 
named as {\em pure-NOMA},  which is the
optimal strategy to achieve the capacity region 
of the considered system. 

\item
We develop design procedures of a {\em hybrid NOMA-OMA} scheme,
where $L$ satellites with \textit{feasible} radio ISLs towards the sink 
are partitioned in groups: namely, those belonging to the same
group implement NOMA, while the ones belonging to different groups
employ OMA. Pure-NOMA and {pure-OMA} techniques are 
special cases of such a scheme, which are 
obtained by setting the number of groups equal to 
$1$ and $L$, respectively.
The objective of the proposed designs is to exploit the differences 
in Doppler frequency profiles of the superimposed 
satellite transmissions in order to 
improve system fairness, by ensuring at the same time a significant
performance gain 
in terms of sum-rate capacity compared to the 
pure-OMA technique.

\item
The performances of the considered multiple access schemes
are validated numerically. It is demonstrated that
exploiting of the inter-plane Doppler frequency diversity is instrumental 
to achieve a good trade-off between sum-rate capacity
and fairness among the satellites.

\end{enumerate}

This article is organized as follows. 
The mathematical model of the considered system and the characterization of 
ISLs feasibility are reported 
in Section~\ref{sec:model}.
The MMSE-SIC receiving structure at the sink is described in 
Section~\ref{sec:NOMA-detection}.
A comparative performance study of the pure-OMA, pure-NOMA, 
and hybrid NOMA-OMA schemes is carried out in 
Section~\ref{sec:capacity}, with particular emphasis on the impact
of the frequency Doppler shifts.
Design procedures of the proposed hybrid NOMA-OMA scheme 
are developed in Section~\ref{sec:opt}, and
related numerical results are reported in Section~\ref{sec:num}. 
Finally, the main
results obtained in this article are summarized and discussed 
in Section~\ref{sec:conc}.

\section{System model}
\label{sec:model}

Let us consider a general LEO Walker Delta constellation, composed of
$K$ satellites, uniformly-distributed
and equally-spaced in $P$ OPs.
Each satellite follows a circular orbit around the Earth,
at a given altitude $h$ above
the terrestrial surface, having an inclination $\alpha$
with respect to the equatorial plane, 
and phasing parameter $F$ \cite{Liang2021}.

Let $N \eqdef K/P$ (assumed to be an integer)
denote the number of satellites in each OP.
In order to univocally identify each satellite,
we use the index pair $(p,n)$ indicating
the $n$-th satellite in the $p$-th orbital plane,
for $n \in \{1,2,\ldots,N\}$ and $p \in \{1,2,\ldots,P\}$.
We investigate the establishment
of ISLs between a number of
transmitting satellites, belonging to a group or
\textit{swarm}, towards a receiving satellite,
referred to as the \textit{sink} in the following.
In particular, without loss of generality,
we choose the satellite $(1,1)$ as the communication sink.
The main parameters and symbols introduced throughout the
paper are summarized in Table~\ref{tab:main}.

\subsection{Conditions ensuring ISL feasibility}
\label{sec: feasibility}

Herein, we investigate the problem
of assuring a reliable
communication link between the 
satellites of each group
and the sink, so called
\textit{ISL feasibility}, which
is influenced by the following aspects:
(i) existence of line-of-sight (LoS) propagation;
(ii) path loss and receiver sensitivity;
(iii) satellite antenna radiation pattern.

To study the first aspect,
we introduce the customary Earth-Centered Earth-Fixed (ECEF)
right-handed orthogonal coordinate system (or \textit{frame}),
with origin at the center of Earth
and $z$ axis coincident with the oriented line
from South pole to North pole
(further details on the definition of the ECEF 
coordinate system are reported in Appendix~\ref{sec:app}).
Assuming the Earth shape to be spherical,
with radius $R = 6378$ km,
the position of the satellite $(p,n)$ at the time instant $t \in \mathbb{R}$,
deployed at altitude $h$ above Earth,
is defined in the ECEF frame by the spherical coordinate triple
$\big(R+h,\theta_{p,n}(t),\varphi_{p,n}(t)\big)$,
where $\theta_{p,n}(t) \in [0, \pi]$ denotes the time-varying polar angle
(or \textit{colatitude}) 
and
$\varphi_{p,n}(t) \in [0,2\pi)$ denotes the time-varying azimuthal angle
(or \textit{longitude}).
Therefore, the distance $d_{p,n}(t)$ at time $t$
between the $(p,n)$-th satellite and the sink
can be calculated with elementary geometry as
\begin{equation}
	d_{p,n}(t) = \|\mathbf{r}_{p,n}(t) - \mathbf{r}_{1,1}(t)\|
\end{equation}
where $\mathbf{r}_{p,n}(t)$ is 
the position in the ECEF coordinate system 
of the $n$-th satellite
belonging to the $p$-th orbit plane
(see Appendix~\ref{sec:app} for its definition).
The corresponding ISL is in LoS visibility if condition 
\be
\textbf{(c1)}: \: 
d_{p,n}(t) \leq d_{\text{horiz}} 
\label{eq:cond1}
\ee
holds, where $d_{\text{horiz}} \eqdef 2 [h \, (h+2 R)]^{1/2}$
is the radio horizon.

Let us consider receiver sensitivity and losses
due to free-space propagation.
For LEO satellites,
propagation occurs mainly in the
\textit{thermosphere}, where the air
molecule density is very low,
hence loss mechanisms due to molecular absorption and scattering
are practically absent \cite{Civas2021}.
When the signal wavelength 
$\lambda_\text{c}$ is very small,
the \textit{geometric path loss} $\big(4 \pi d_{p,n}(t)/\lambda_{\text{c}}\big)^2$
is considerable, so highly-directional
antennas with high gains are required.
Therefore, we assume that each satellite is equipped
with highly-directional wideband antennas,
with approximately conical beams,
whose half-beamwidths are denoted with
$\beta_{p,n}$
(assumed to be equal for all the antennas on board each satellite).
In this case, assuming a free-space path loss model, the received power is given
by the Friis formula (see, e.g., \cite{Tse-book})
\be
\text{P}_{p,n}^{\text{rx}}(t) = \text{P}_{p,n}^{\text{tx}} \, \text{G}_{p,n}^{\text{tx}} \, \text{G}_{1,1}^{\text{rx}}  \left[ \frac{\lambda_\text{c}}{4 \pi d_{p,n}(t)} \right]^2
\label{eq:power}
\ee
where $\text{P}_{p,n}^{\text{tx}}$ and
$\text{P}_{p,n}^{\text{rx}}(t)$ are
the power transmitted by the $(p,n)$-th satellite
and received at the sink at time $t$, respectively,
$\text{G}_{p,n}^{\text{tx}}$ and $\text{G}_{1,1}^{\text{rx}}$
represent the antenna gains of the $(p,n)$-th transmitting
satellite and sink.\footnote{In general, the antenna gains
depends on $t$ since the propagation direction
is time-varying. However, in the case of 
highly-directional antennas, the gains can 
be reasonably assumed to be constant
on a spherical cap.}
 
For the  propagation link between
the $(p,n)$-th transmitting satellite  and the sink, 
the power $\text{P}^{\text{rx}}_{p,n}(t)$ has to fulfill the condition
\be
\textbf{(c2)}:
\text{P}^{\text{rx}}_{p,n}(t) \ge \text{P}^{\text{rx}}_{\text{sens}}
\label{eq:cond3}
\ee
where the receiver sensitivity $\text{P}^{\text{rx}}_{\text{sens}}$
represents
the lowest value of the received
power at which the signal
can be decoded satisfactorily,
i.e., with a given bit-error-rate (BER).

\begin{figure}
	\centering
	\includegraphics[width=0.8\columnwidth]{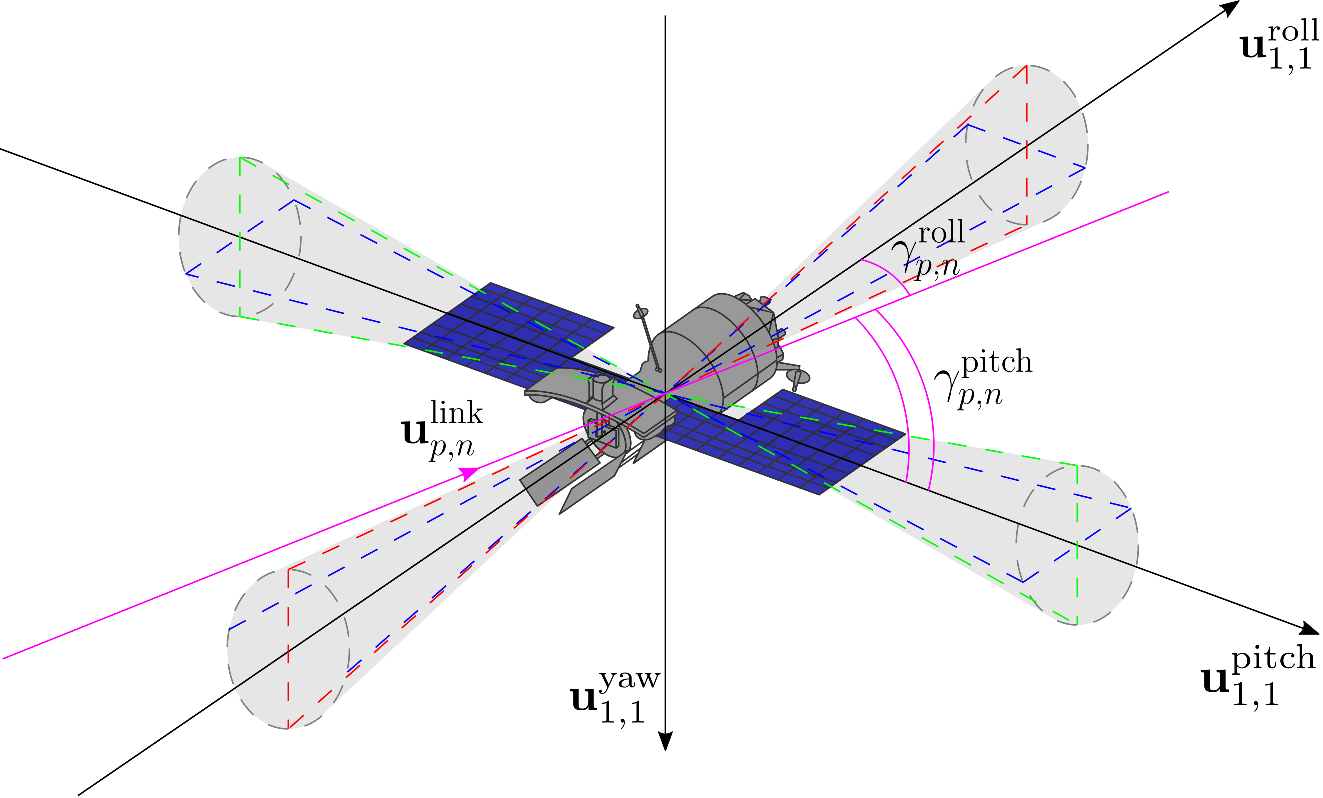}
	\caption{The ISL antenna beams of the sink (the time variable $t$ is omitted).}
	\label{fig:fig_1}
\end{figure}

Finally, we take into account antenna radiation
pattern limitations.
Usually, a spacecraft is equipped with intra-plane ISL antennas
located at both sides of the roll axis
(aligned to the satellite speed vector)
and inter-plane ISL antennas along
the pitch axis (perpendicular to
the orbit plane),
as shown in Fig. \ref{fig:fig_1}.
Hereinafter, we assume that each satellite is equipped with antennas having approximately conical beams, 
whose half-beamwidths $\beta_{p,n}$ are equal for all the antennas on board each satellite.
Consequently, we assume that a feasible LoS link
between satellite $(p,n) \ne (1,1)$
and the sink $(1,1)$ can be established
only if the corresponding RF propagation
direction falls within the conical antenna beams of the sink.
Mathematically, let $\gamma^{\text{pitch}}_{p,n}(t)$  and
$\gamma^{\text{roll}}_{p,n}(t)$ denote the time-varying angles
between ${\bf u}^{\text{link}}_{p,n}(t)$, i.e.,
the direction of propagation of the signal emitted by
the $(p,n)$-th satellite, and the pitch and roll axes
of the sink, given respectively by ${\bf u}^{\text{pitch}}_{1,1}(t)$
and ${\bf u}^{\text{roll}}_{1,1}(t)$ (see Fig. \ref{fig:fig_1}),
the condition must hold
\be
\textbf{(c3)}:
\: \{\gamma^{\text{pitch}}_{p,n}(t) \le \beta_{1,1}\} \: \cup \: \{\gamma^{\text{roll}}_{p,n}(t) 
\le \beta_{1,1}\}\:  \,\,
\text{for $(p,n) \ne (1,1)$}.
\label{eq:cond2}
\ee

\begin{definition}[{\em Feasible ISLs}]
\label{def:feasible}
{\em 
Only those satellites
that satisfy the \textit{three}
ISL feasibility requirements \textbf{(c1)}, \textbf{(c2)}, and \textbf{(c3)},
can effectively communicate with the sink.
The time-varying set $\mathcal{C}(t)$
contains all the couples
$(p,n)$ satisfying \eqref{eq:cond1},
\eqref{eq:cond2}, \eqref{eq:cond3} and
$L(t) \eqdef \card\big(\mathcal{C}(t)\big) \le K$ is the number
of feasible ISLs.
}
\end{definition}

\subsubsection{Illustrative example}
\label{sec:example}
\begin{figure}
\centering
\includegraphics[width=0.8\columnwidth]{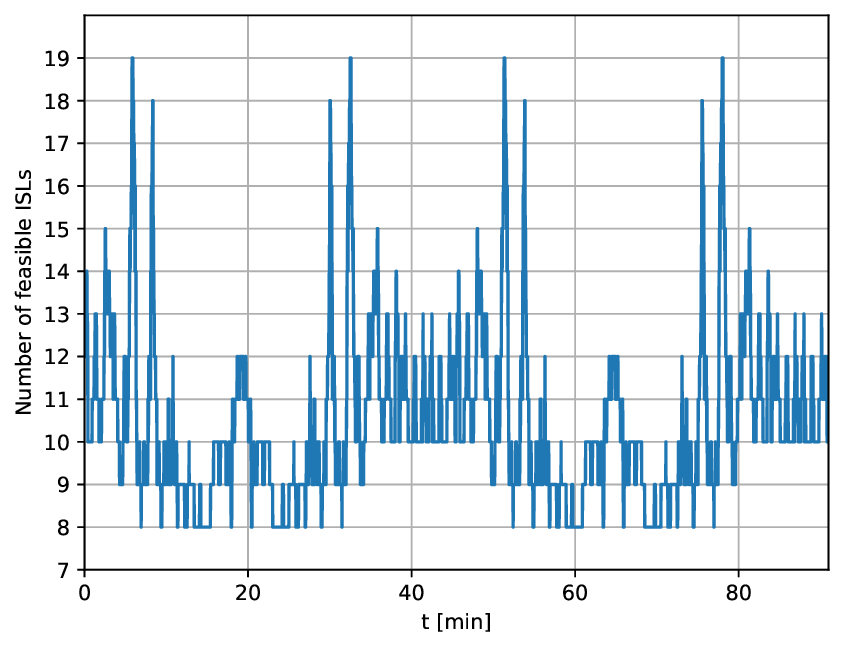}
\caption{Number of feasible ISLs over the time-scale  
$T_\text{rev}$.
}
\label{fig:fig_2}
\end{figure}

We consider a LEO Walker Delta constellation 
composed of $K=1584$ satellites flying over $P=22$ OPs at altitude $h=550$ km.
The number of satellites per OP is $N=K/P=72$,
the inclination with respect to the equatorial plane is $\alpha=53^\circ$,
and the phasing parameter (see Appendix~\ref{sec:app}) is set equal to $F=17$.
Moreover, the period of revolution $T_\text{rev}$ of the satellites is $91$ minutes.
We numerically evaluate in Fig.~\ref{fig:fig_2}
the total number $L(t)$ of feasible ISLs
as a function of $t$, when the sink corresponds
to the satellite having indices $(15,47)$,
over a time horizon equal to $T_\text{rev}$.
In this example, the signal carrier 
frequency $f_\text{c}=c/\lambda_\text{c}$ is $40$ GHz, 
with $c$ denoting the speed of light in the vacuum, 
the power transmitted by the satellites is $\text{P}_{p,n}^{\text{tx}}=10$ W, 
the transmit and receive antenna gains are set
equal to $20$ dBi,\footnote{Considering the assumptions on the antenna
radiation patterns, a gain of $20$ dBi roughly
corresponds to 11.48 degrees half--beamwidth.}
the receiver sensitivity is $\text{P}^{\text{rx}}_{\text{sens}}=-120$ dBm.
It is interesting to note that the number of feasible ISLs
seen by the sink over  the time-scale $T_\text{rev}$ exhibits a 
staircase trajectory, i.e., $L(t)$ remains constant over time intervals
of  length $\Delta_{i,\pi(i)}$, with $i \in \{8,9, \ldots, 19\}$ 
denoting the total number of feasible ISLs, i.e., $L(t)=i$, and 
$\pi(i) \in \mathbb{N}$ being the index of the time interval
within $L(t)$ takes on the value $i$.
Moreover, potential values of $L(t)$ repeat over the revolution time 
horizon, e.g., the value $L(t)=19$ is taken on four times
over  the time-scale $T_\text{rev}$ and, thus, 
$\pi(19) \in \{1,2,3,4\}$.
The minimum value of $L(t)$ is equal to $8$, which corresponds to the
(fixed) number of feasible intra-OP ISLs seen by the sink over each revolution 
time interval. Therefore, the number of feasible inter-OP ISLs varies from 
$1$ (i.e., $i=9)$ to $11$ (i.e., $i=19$).
It should be observed that, due to the periodicity of the orbits,
the same result can be extended to all of the satellites
of the constellation by a simple cyclic shift of the plot.

\begin{figure}
\centering
\includegraphics[width=0.8\columnwidth]{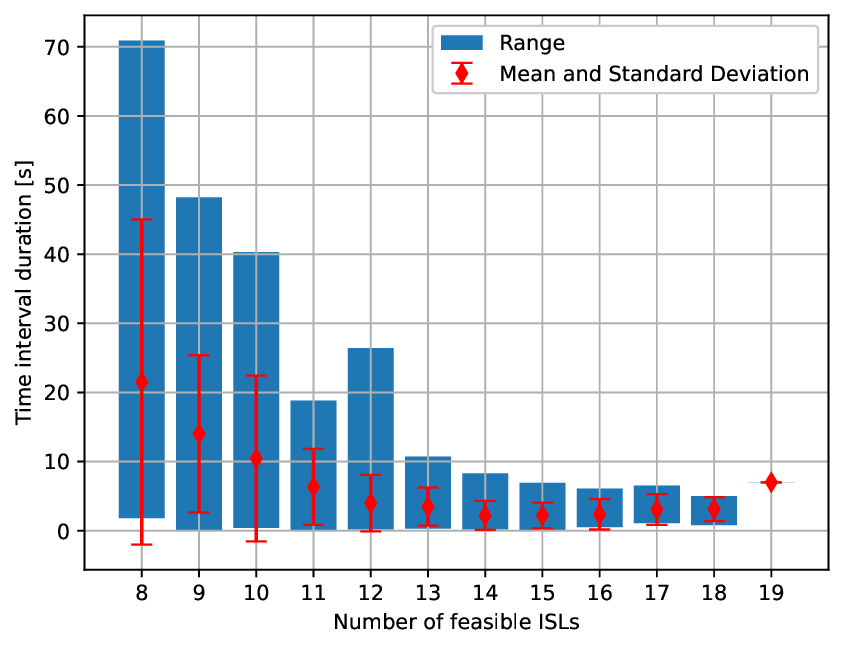}
\caption{Range, mean value, and standard deviation of 
the time interval duration $\Delta_{i,\pi(i)}$
versus the number $i$ of feasible ISLs over the time-scale 
$T_\text{rev}$.}
\label{fig:fig_3}
\end{figure}

\begin{figure}
\centering
\includegraphics[width=0.8\columnwidth]{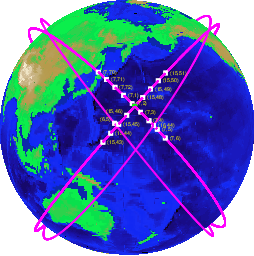}
\caption{A pictorial view of the $i=19$ satellites with feasible ISLs
towards the sink represented by the satellite with indices $(15,47)$ 
(i.e., the green square). 
}
\label{fig:fig_4}
\end{figure}

For each value of  $L(t)=i$, we 
also show in Fig.~\ref{fig:fig_3} 
the range $[\Delta_{i,\text{min}},\Delta_{i,\text{max}}]$ of $\Delta_{i,\pi(i)}$, 
where $\Delta_{i,\text{min}} \eqdef \min_{\pi(i)}\Delta_{i,\pi(i)}$
and $\Delta_{i,\text{max}} \eqdef \max_{\pi(i)}\Delta_{i,\pi(i)}$,
as well as its (arithmetic) mean value 
and the corresponding standard deviation. 
It is worthwhile to note that, besides $i=8$ representing the 
number of feasible intra-OP ISLs, there exist multiple inter-OP ISLs 
towards the sink within time windows of non-negligible length.
For instance, the sink sees $i=19$ feasible ($8$ intra-OP plus
$11$ inter-OP) ISLs over four time windows having 
(approximately) the same length equal to $7.025$ s.
A snapshot of this special case is depicted in Fig.~\ref{fig:fig_4}. 

Hereinafter, for the sake of notational simplicity,
we refer to a generic time window 
$[0, T_0)$ of length $T_0$, referred to
as  the {\em observation interval}, 
over which  
$\mathcal{C}(t)$ is constant with  $L(t) \equiv L$, thus
omitting the dependence of both $\mathcal{C}(t)$
and $L(t)$ on $t$. 
Moreover, 
it is assumed that a suitable 
mapping has been introduced, which transforms each
two-dimensional (2D) coordinate $(p,n) \in \mathcal{C}$
into a one-dimensional (1D) index $\ell \in
\mathcal{L} \eqdef \{1, 2, \ldots, L\}$.

\subsection{The proposed multiple-access transmission scheme}
\label{sub:access}

We briefly introduce herewith the proposed
multiple-access scheme (further details are given in 
Sections~\ref{sec:NOMA-detection} and 
\ref{sec:capacity}).
For a given value of $G \in \{1,2,\ldots, L\}$, let 
$\mathbb{P}({\mathcal{L}}) = \{ \mathcal{L}_k\}_{k=1}^{G}$ 
be a partition of $\mathcal{L}$,
i.e., $\cup_{k=1}^G \mathcal{L}_k= \mathcal{L}$ and 
$\mathcal{L}_k \cap \mathcal{L}_h =\emptyset$ for $k \neq h$,
with $\sum_{k=1}^G L_k=L$, where
$L_k \eqdef \card(\mathcal{L}_k)$. 
Our aim is to determine $\mathbb{P}({\mathcal{L}})$
such that the satellites that are members of the group $\mathcal{L}_k$
concurrently transmit by employing a NOMA
scheme, 
based on superposition coding (SC)
and SIC 
at the receiver, 
whereas the satellites belonging to different
groups $\mathcal{L}_k$ and $\mathcal{L}_h$, with $k \neq h$, 
transmit by employing an OMA scheme over
non-interfering links. 
It is noteworthy that it is irrelevant for 
a capacity analysis whether the partitioning is across
time or frequency (or both), since 
the power constraint is on the average across
the degrees of freedom (DoF) \cite{Tse-book}.

We would like to recall attention to 
two extreme cases: when $G=1$ and, thus, $L_1=L$,
all the $L$ satellites
with feasible links transmit over 
the same time-frequency resources 
by using the SC-SIC scheme and, in this case, 
one has a {\em pure-NOMA} technique. 
On the other hand, when $G=L$ and, thus, $L_k=1$, 
the sink receives
orthogonal signals from the $L$ different satellites  
and, in this case, the proposed scheme boils down to 
a {\em pure-OMA} technique.
For the intermediate values of 
$G \in \{2, 3, \ldots, L-1\}$, the satellites communicate
with the sink through a {\em hybrid NOMA-OMA} transmission scheme.


\subsection{Model of the signal received by the group $\mathcal{L}_k$}
\label{sec:rec-signal}

We assume that
the partition $\mathbb{P}(\mathcal{L})$
has been determined and,
without loss of generality,
we focus on NOMA communication
within the $k$-th satellite
group $\mathcal{L}_k \eqdef \{\ell_{k,1}, \ell_{k,2}, \ldots, \ell_{k,L_k}\}$, 
for $k \in \{1,2,\dots, G\}$.
Moreover, let $T$ be the symbol period,
$\{b_{{\ell_{k,q}}}(h) \}\in \Cset$ the sequence
sent by the $({\ell_{k,q}})$-th satellite
in the $h$-th symbol interval, with $h \in \mathbb{Z}$
and $q \in \{1,2,\ldots,L_k\}$, 
and let $\psi(t)$ denote the impulse response
of the transmit shaping filter.
With reference to a linear modulation, the baseband signal
of the $({\ell_{k,q}})$-th satellite passed through the transmit shaping filter
is given by
\be
x_{\ell_{k,q}}(t) = \frac{1}{\sqrt{\rho_k}}\sum_{h=-\infty}^{+\infty} b_{_{\ell_{k,q}}}(h) \, \psi(t-h\,T) \:,
\quad \text{for $t \in [0, T_0)$}
\ee
where 
$\rho_{k}$ is the fraction of DoF allocated
to all the satellites belonging to the group $\mathcal{L}_k$, with 
\be
\sum_{k=1}^{G} \rho_{k} =1 \: .
\label{eq:rhok}
\ee
The transmitted symbols $\{b_{_{\ell_{k,q}}}(h)\}_{h \in \mathbb{Z}}$ are modeled as mutually
independent sequences of zero-mean unit-variance independent and identically distributed (i.i.d.)
random variables.
Hereinafter, we assume that the $({\ell_{k,q}})$-th
satellite is moving (relative to the sink)
at constant \textit{radial} speed 
$v_{\ell_{k,q}}$
within the observation interval. 
In this case, the time-varying distance covered by the wavefront transmitted 
by the satellite $\ell_{k,q}$ and received at the sink is given by 
$d_{\ell_{k,q}}(t)=d_{\ell_{k,q}}(0) +v_{\ell_{k,q}} \, t$, 
for $t \in [0, T_0)$.
Under the assumption that $|v_{\ell_{k,q}}| \, T_0 \ll |d_{\ell_{k,q}}(0)|$, 
the Doppler effect induces a simple carrier
frequency shift \cite{Napolitano2014} and, thus, 
the far-field noise-free signal $r_k(t)$ received by the sink from the $k$-th satellite group can 
be expressed as
\be
r_k(t) = \sum_{q=1}^{L_k} A_{\ell_{k,q}} \, e^{j 2 \pi f_{\ell_{k,q}} t} \, x_{\ell_{k,q}}
(t-\tau_{\ell_{k,q}}) 
\label{eq:sig_ric}
\ee
for $t \in [0, T_0)$, where $A_{\ell_{k,q}}$, $f_{\ell_{k,q}}$,
and $\tau_{\ell_{k,q}}$ are the
complex amplitude, Doppler frequency shift, and delay, respectively,
of the link between the
$(\ell_{k,q})$-th satellite and the sink.
Model \eqref{eq:sig_ric} has been derived under the
additional assumption that $f_{k,\text{max}} \, T \ll 1$,
where $f_{k,\text{max}} \eqdef \text{max}_{q \in\{1,2,\ldots,L_k\}} |f_{\ell_{k,q}}|$ is the
maximum Doppler frequency shift in $\mathcal{L}_k$.
It should be noted that the argument of  $A_{\ell_{k,q}}$ accounts for
a phase misalignment between the 
$(\ell_{k,q})$-th satellite and the sink.

In the following, we will assume that a timing advance (TA)
scheme is employed \cite{3gpp.38.811}, such that
the signals transmitted by satellites
within the same time interval arrive
nearly aligned at the sink. In this case, 
$\tau_{\ell_{k,q}} \approx \tau_0$
in \eqref{eq:sig_ric} and 
timing misalignment interference is avoided.
Without loss of generality, we set hereinafter $\tau_0 = 0 $ to simplify the
forthcoming derivations.

At the sink, to demodulate the symbols
received in the interval $[u \,T, (u+1) \,T] \subseteq [0, T_0)$ ($u \in \Zset$), 
the signal \eqref{eq:sig_ric} is first passed through 
a receiving filter that is matched to the transmit pulse shaping characteristic
and, then, sampled
with rate $1/T_{\text{c}} \eqdef S/T$ at instants
$t_{u,s} \eqdef u \,T + s \, T_{\text{c}}$,
for $s \in \{0,1,\ldots, S-1\}$, with
$S > 1$ denoting the
oversampling factor.
Let $r_{k}^{(s)}[u] \eqdef r_k(t_{u,s})$ be the
discrete-time version of \eqref{eq:sig_ric}, 
for $u \in \{0, 1, \ldots, N_0-1\}$, one gets
\be
	r_{k}^{(s)}[u] = \sum_{q=1}^{L_k}  \frac{A_{\ell_{k,q}}}{\sqrt{\rho_k}} \, e^{j  2 \pi S \nu_{\ell_{k,q}} u} \, e^{j  2 \pi \nu_{\ell_{k,q}} s} \, 
	             p(s \, T_{\text{c}}) \, b_{\ell_{k,q}}[u]  
\label{eq:modellost}
\ee
with $N_0 \eqdef T_0/T$ (assumed to be an integer for simplicity), 
the composite pulse $p(t) \eqdef \psi(t) \ast \psi(T-t) \approx 0$ 
$\forall t \not \in (0, 2 \, T)$,\footnote{We have chosen a pulse shaping of 
finite time duration to maintain models and analysis at a reasonably 
simple level. Our framework can be generalized by 
considering other types of pulse shaping filters, e.g., 
the raised-cosine one.}
and 
$\nu_{\ell_{k,q}} \eqdef f_{\ell_{k,q}} T_{\text{c}} \in [0,1)$
denoting the \textit{normalized} Doppler frequency shift.
Model~\eqref{eq:modellost} is valid under the further assumption that 
Doppler fluctuations vary at a very slow pace compared to the effective duration
of the receiving filter. 

Let $\bm{r}_k[u] \eqdef \left[r_{k}^{(0)}[u], 
r_{k}^{(1)}[u],\ldots, r_{k}^{(S-1)}[u]\right]^{\trasp} \in \Cset^S$,
it results that, for $u \in \{0, 1, \ldots, N_0-1\}$, 
\barr
	\bm{r}_k[u] &= \sum_{q=1}^{L_k}  \frac{A_{\ell_{k,q}}}{\sqrt{\rho_k}} \, e^{j 2 \pi S \nu_{\ell_{k,q}} u} \, \widetilde{\bm P} \,
	\bm{v}_{\ell_{k,q}} \, b_{\ell_{k,q}}[u]  \nonumber \\
	&= \frac{1}{\sqrt{\rho_k}}\sum_{q=1}^{L_k}  \widetilde{\bm h}_{\ell_{k,q}}[u] 
	\, b_{\ell_{k,q}}[u] 
	\label{eq:vec_ric}
\earr
where $\widetilde{\bm{h}}_{\ell_{k,q}}[u] \eqdef A_{\ell_{k,q}} \, e^{j 2 \pi S \nu_{\ell_{k,q}} u}
\, \widetilde{\bm{P}} \, \bm{v}_{\ell_{k,q}} \in \Cset^S$ represents the  {\em composite} channel response
of the $(\ell_{k,q})$-th ISL, with
\barr
\widetilde{\bm{P}} & \eqdef \text{diag}[p(0), p(T_{\text{c}} ), \ldots, p((S-1) \, T_{\text{c}})]
\\
\bm v_{\ell_{k,q}} & \eqdef [1,e^{j 2 \pi \nu_{\ell_{k,q}}}, \ldots, e^{j 2 \pi \nu_{\ell_{k,q}} (S-1)}]^{\trasp} \in \Cset^{S} \:.
\earr
It should be observed that $\bm v_{\ell_{k,q}}$ is 
a Vandermonde vector.
In the sequel, we assume that $\widetilde{\bm{P}}$ is non-singular.

\section{Signal detection}
\label{sec:NOMA-detection}

In this section, we describe the reception
strategy adopted in the proposed NOMA scheme.
We assume that the sink is perfectly aware of all channel parameters in
\eqref{eq:vec_ric}, i.e., complex amplitudes $A_{\ell_{k,q}}$ and
normalized Doppler frequency shifts $\nu_{\ell_{k,q}}$,
which can be calculated from the knowledge of the 
deterministic motion of the satellites by means of simple geometrical arguments.
This knowledge is exploited at the sink
to perform simultaneous satellite symbol detection.

The overall signal received by the sink can be written as
\be
\bm{r}[u] = \sum_{k=1}^G \bm{r}_k[u] + \bm w[u]\:,
\quad \text{for $u \in \{0, 1, \ldots, N_0-1\}$}
\label{eq:r-sig}
\ee
where $\bm w[u] \eqdef \big[w^{(0)}[u],w^{(1)}[u],\ldots,
w^{(S-1)}[u]\big]^{\trasp} \in \Cset^S$,
with $w^{(s)}[u] \eqdef w(t_{u,s})$, and  
$w(t)$ represents the complex envelope of noise,
modeled as a circularly-symmetric complex Gaussian
random process with statistical correlation function $c_{ww}(\tau) \eqdef \Es[w(t) \, w^*(t-\tau)] = \sigma_w^2 \, p(\tau+T)$, which is
assumed to be independent of $x_{\ell_{k,q}}(t)$,
for any $\ell_{k,q} \in \mathcal{L}_k$ and 
$k \in \{1,2,\ldots, G\}$.
The positive definite correlation matrix of $\bm w[u]$ can be expressed as 
$\bm C_{\bm w \bm w} \eqdef \Es(\bm w[u] \, \bm w^\herm[u]) =
\sigma^2_w \, \bm C_{\mathrm{pp}}$, where
the $(s_1+1,s_2+1)$-th entry of $\bm C_{\mathrm{pp}}$ 
is given by 
\be
\left\{\bm C_{\mathrm{pp}}\right\}_{s_1+1,s_2+1} = p\left((s_1-s_2) \, T_{\text{c}}+T\right)
\ee
for $s_1, s_2 \in \{0,1,\ldots, S-1\}$.
Since $\bm C_{\mathrm{pp}}$ depends only on the pulse shaping 
waveform, it is perfectly known at the sink and, thus, the colored 
noise can be whitened with a whitening filter.
Cholesky factorization determines the invertible whitening transformation according
to $\bm C_{\mathrm{pp}}=\bm T_{\mathrm{pp}} \, \bm T_{\mathrm{pp}}^\trasp$, where 
$\bm T_{\mathrm{pp}} \in \Rset^{S \times S}$
is a lower triangular matrix.
To whiten $\bm w[u]$, the sink passes $\bm{r}[u]$ 
through the matrix multiply $\bm T_{\mathrm{pp}}^{-1}$.
By the theorem of reversibility \cite{Proakis2007}, no information is lost in such a transformation.
At this point, it is convenient to build  a more compact vector model.
In particular, let us denote with
$\bm{b}_k[u] \eqdef \big[b_{\ell_{k,1}}[u], b_{\ell_{k,2}}[u], \ldots, b_{\ell_{k,L_k}}[u]\big]^{\trasp} \in \Cset^{L_k}$ 
the block collecting the symbols transmitted by all satellites
in $\mathcal{L}_k$,
one has from \eqref{eq:vec_ric} and \eqref{eq:r-sig} that, at the output of the whitening filter,  
the received signal reads as 
\be
\overline{\bm{r}}[u] \eqdef \bm T_{\mathrm{pp}}^{-1} \, \bm{r}[u] 
= \sum_{k=1}^G \widetilde{\bm{H}}_k[u] \, \bm{b}_k[u] + \overline{\bm{w}}[u]
\label{eq:block}
\ee
for $u \in \{0, 1, \ldots, N_0-1\}$, where 
\barr
\widetilde{\bm{H}}_k[u] & \eqdef 
\frac{1}{\sqrt{\rho_k}}\left[\widetilde{\bm h}_{\ell_{k,1}}[u], \widetilde{\bm h}_{\ell_{k,2}}[u], \ldots,
\widetilde{\bm h}_{\ell_{k,L_k}}[u]\right] 
\nonumber \\ & = \frac{1}{\sqrt{\rho_k}} \, \bm{P} \, \bm{V}_k \, \bm{A}_k \, \bm{E}_k[u]
\in \Cset^{S \times L_k}
\label{eq:mat_H_tilde}
\earr
with $\bm{P} \eqdef \bm T_{\mathrm{pp}}^{-1} \, \widetilde{\bm{P}} \in \Cset^{S \times S}$
positive definite,
$\bm{V}_k \eqdef  [\bm{v}_{\ell_{k,1}}, \bm{v}_{\ell_{k,2}}, \ldots, \bm{v}_{\ell_{k,L_k}}] \in\Cset^{S \times L_k}$ 
being a Vandermonde matrix, whose rank properties will be
discussed soon after in Subsection~\ref{sec:discuss}, 
whereas the diagonal matrices
$\bm{A}_k  \eqdef \text{diag}[A_{\ell_{k,1}}, A_{\ell_{k,2}}, \ldots, A_{\ell_{k,L_k}}]$
and  
$\bm{E}_k[u]  \eqdef  \text{diag}[e^{j 2 \pi S \nu_{\ell_{k,1}} u}, e^{j 2 \pi S \nu_{\ell_{k,2}} u}, \ldots, e^{j 2 \pi S \nu_{k,\ell_{L_k}} u}] $
are non-singular, and, finally,  
$\overline{\bm{w}}[u] \eqdef \bm T_{\mathrm{pp}}^{-1} \, \bm{w}[u]$.

We recall that the satellites belonging to different
groups $\mathcal{L}_k$ and $\mathcal{L}_h$, with $k \neq h$, 
transmit over point-to-point orthogonal links, which have been
assigned at a system level by means of a deterministic 
{\em channel mapping} procedure. Consequently, 
the contribution $\overline{\bm{r}}_k[u]$
of the $k$-th group can be perfectly recovered from 
\eqref{eq:block} through a  
{\em channel demapping} operation, thus obtaining,
for $u \in \{0, 1, \ldots, N_0-1\}$, 
\be
\overline{\bm{r}}_k[u] = 
\widetilde{\bm{H}}_k[u] \, \bm{b}_k[u] + \overline{\bm w}_k[u]
\label{eq:rbar}
\ee
where $\overline{\bm w}_k[u]$ is noise after the 
channel demapping algorithm, which is
a circularly-symmetric zero-mean 
complex Gaussian random vector 
with covariance matrix $\sigma_{w}^2 \, \bm I_{S}$.

For each symbol period and satellite group, the detection algorithm consists
of an iterative procedure with $L_k$ iterations, 
which sequentially decodes each entry of $\bm b_k[u]$,
by repeating the following three steps per each iteration:
(i) MMSE filtering;
(ii) maximum signal-to-interference-plus-noise ratio (SINR) ISL selection;
(iii) SIC.

In Subsections~\ref{sec:MMSE}, \ref{sec:max-SINR} and \ref{sec:SIC}, since 
we focus on the $u$-th symbol
interval and $k$-th group $\mathcal{L}_k$, we omit the dependence on both $u$ and $k$ for notational convenience (we maintain the subscript $k$ only for the cardinality $L_k$).


\subsection{MMSE filtering}
\label{sec:MMSE}

Let $\overline{\bm r}^{(m)} \in \Cset^{S}$ denote the input vector of 
the MMSE filter at the $m$-th iteration, for $m \in \{0,1,\ldots,L_k-1\}$, where
$\overline{\bm{r}}^{(0)} \equiv \overline{\bm{r}}$
is given by \eqref{eq:rbar}.
As a first step, the receiver performs MMSE
filtering of $\overline{\bm r}^{(m)}$ as follows
\be
\bm z^{(m)} \eqdef [z^{(m)}_{1}, z^{(m)}_{2},\ldots,z^{(m)}_{L_k-m}]^\trasp
= \widetilde{\bm F}^{(m)} \, \overline{\bm r}^{(m)}
\ee
which represents a soft estimate of
$\bm b^{(m)} \in \Cset^{L_k-m}$, with $\bm b^{(0)} \equiv \bm b$,
and $\widetilde{\bm F}^{(m)} \in \Cset^{(L_k-m) \times S}$
is the solution of the MMSE problem
$\min_{\widetilde{\bm F}^{(m)}} \: \mathbb{E}[\| \widetilde{\bm F}^{(m)} 
\, \overline{\bm r}^{(m)} - \bm b^{(m)} \|^2]$,
whose solution is given by
\be
\widetilde{\bm F}^{(m)}  = \{\widetilde{\bm H}^{(m)}\}^\herm 
\left[\widetilde{\bm H}^{(m)} \,  
\{\widetilde{\bm H}^{(m)}\}^\herm  + \sigma_w^2 \, \bm I_{S}\right]^{-1}
\label{eq:mmse}
\ee
where $\widetilde{\bm H}^{(m)} \in \Cset^{S \times (L_k-m)}$, 
with  $\widetilde{\bm H}^{(0)} \equiv \widetilde{\bm H}$.
The quantities  $\overline{\bm r}^{(m)}$, 
$\widetilde{\bm H}^{(m)}$, and $\bm b^{(m)}$ 
iteratively depend on  $\overline{\bm r}^{(m-1)}$,  $\widetilde{\bm H}^{(m-1)}$, and $\bm b^{(m-1)}$, 
respectively, as detailed in Subsection \ref{sec:SIC}.

\subsection{Maximum SINR ISL selection}
\label{sec:max-SINR}

The maximum SINR (max-SINR) ISL selection algorithm is employed to identify, at each iteration,
the ``best'' ISL, i.e., the entry of the vector $\bm z^{(m)}$ that exhibits the
highest SINR at the output of the MMSE filter.
To this aim, at the $m$-th iteration, the SINR
on the $\ell$-th entry of $\bm z^{(m)}$ is given \cite{Li2006} by
\be
	\text{SINR}^{(m)}_{\ell} 
	= \frac{1}{\sigma_w^2 \left[ \left(\sigma_w^2 \, \bm I_{L_k-m} +
		\{\widetilde{\bm H}^{(m)}\}^\herm \, \widetilde{\bm H}^{(m)} \right )^{-1}\right]_{\ell\ell}}-1 
	\label{eq:sinr-ex}
\ee
with $\ell \in \{1,2,\ldots, L_k-m\}$.
Hence, at iteration $m$, we seek the index $\ell_{\text{max}}^{(m)}$ that solves
\be
\ell_{\text{max}}^{(m)} = \arg \max_{\ell \in \{1,2,\ldots,L_k-m\}} \text{SINR}^{(m)}_{\ell}
\ee
and, thus, we get the hard estimate 
$\widehat{b}_{{\ell_{\text{max}}^{(m)}}} 
\eqdef \mathcal{Q}[z^{(m)}_{\ell_{\text{max}}^{(m)}}]$
of $b_{{\ell_{\text{max}}^{(m)}}}$, 
where $\mathcal{Q}[\cdot]$ is the minimum-distance decision metric,
which depends on the constellation alphabet.


\subsection{Successive interference cancellation}
\label{sec:SIC}

The final step of the iterative decoding algorithm
consists of cancelling out, at each iteration,
the contribution of $\widehat{b}_{{\ell_{\text{max}}^{(m)}}}$
to the remaining symbols, as
\be
\overline{\bm r}^{(m+1)} = \overline{\bm r}^{(m)} -  \widehat{b}_{{\ell_{\text{max}}^{(m)}}} \,
\widetilde{\bm h}_{\ell_{\text{max}}^{(m)}}^{(m)}
\ee
where $\widetilde{\bm h}_{\ell_{\text{max}}^{(m)}}^{(m)}$ is the $[\ell_{\text{max}}^{(m)}]$-th column of $\widetilde{\bm H}^{(m)}$.
Under the simplifying assumption that
$\widehat{b}_{\ell_{\text{max}}^{(m)}} \equiv b_{\ell_{\text{max}}^{(m)}}$
(correct decision),
the received vector at the $(m+1)$-th
iteration becomes
\be
\overline{\bm{r}}^{(m+1)} = \widetilde{\bm{H}}^{(m+1)} \, \bm{b}^{(m+1)} + \overline{\bm{w}}
\label{eq:rmone}
\ee
where the matrix $\widetilde{\bm H}^{(m+1)}$ is obtained from
$\widetilde{\bm H}^{(m)}$ by removing its $[\ell_{\text{max}}^{(m)}]$-th column and, similarly,
the vector $\bm b^{(m+1)} $ is obtained from $\bm b^{(m)}$ by removing its
$[\ell_{\text{max}}^{(m)}]$-th entry.

\subsection{Remarks}
\label{sec:discuss}

The assumption of correct decisions is often invoked when a theoretical capacity  
analysis is of concern, but of course errors may be made during the SIC process:
if one satellite is decoded incorrectly,
all subsequent satellites in the same group are affected
(so-called {\em error propagation} phenomenon).
If all the satellites  are coded with the same target error probability
assuming error-free SIC, the effect of error propagation degrades the
error probability by a factor of at most the number of 
satellites $L_k$ \cite{Tse-book}. This fact drives one to choose
$L_k$ reasonably small such that error propagation effects 
can be compensated by slightly increasing the 
coding block length.

As a second remark, we underline that the sum rate of the 
$L_k$ satellites belonging to the $k$-th  group $\mathcal{L}_k$
is independent of the satellite ordering and 
achieves the boundary
of the multiple-access channel (MAC) capacity region 
(see Section~\ref{sec:capacity} for details).
However, the individual rates are affected by the satellite ordering: indeed, 
satellites decoded at
a later stage of the SIC process can benefit from increased throughput. 
The maximum-SINR ordering described in 
Subsection~\ref{sec:max-SINR} guarantees that
satellites with better channel conditions are decoded earlier,
thus allowing subsequently-decoded weaker satellites to get 
the best possible rate. Such a sharing strategy ensures
{\em max-min fairness} \cite{Ber-book}. 

A third remark is in order regarding the rank property of 
$\widetilde{\bm{H}}_k[u]$ given by \eqref{eq:mat_H_tilde}. 
Since the matrices 
$\bm{P}$, $\bm{A}_k$, and $\bm{E}_k[u]$ are non-singular, one has  
$\text{rank}(\widetilde{\bm{H}}_k[u])=\text{rank}(\bm{V}_k)$. 
Relying on the properties of Vandermonde vectors \cite{Horn2012},
one gets $\text{rank}(\bm{V}_k)=\min(S, L_k)$,
provided that 
\be
\textbf{(c4)}:
\: 
\nu_{\ell_{k,1}} \neq \nu_{\ell_{k,2}} \neq \cdots \neq \nu_{\ell_{k,L_k}} \: .
\label{eq:cond4}
\ee
Condition \textbf{(c4)} states that the Doppler frequency shifts of the satellites 
belonging to the $k$-th group $\mathcal{L}_k$ have to be (significantly) different
from each other in order to obtain a (well-conditioned) full-rank channel matrix.
It is worth noticing that not all satellites with feasible ISLs towards the sink 
(see Definition~\ref{def:feasible}) have different 
Doppler frequency shifts. 
To highlight such a fact, we have reported in Table~\ref{tab:doppler}
the Doppler frequency shifts of the $L=19$ satellites of Fig.~\ref{fig:fig_4} 
(see also Subsection~\ref{sec:example}) with
feasible ISLs towards the sink $(15,47)$ in the observation interval of 
duration $T_0=0.5$ ms (corresponding
to the transmission of $N_0=2000$ symbols at the rate of $1/T=4$ Mbaud).  
It is seen from Table~\ref{tab:doppler} that all the eight satellites belonging 
to the same OP of the sink (i.e., those with $p=15$)
essentially exhibit the same Doppler frequency shift, which is 
very close to zero. 
However, the satellites that do not belong to the $15$-th OP 
experience high Doppler frequency shifts, especially those
flying over the crossing OP $7$.
As we will show in Section~\ref{sec:capacity}, 
this behavior has a crucial impact on 
the condition number of the channel matrix 
$\widetilde{\bm{H}}_k[u]$ and, thus, on 
the design of the groups. 

\begin{table}
\caption{Doppler frequency shifts of the satellites with feasible ISLs 
towards the sink $(15,47)$.}
\label{tab:doppler}
\centering{}%
\begin{tabular}{ccc}
\hline
\noalign{\vskip\doublerulesep}
\textbf{2D index} & \textbf{1D index} & \textbf{Doppler frequency shift (Hz)} \tabularnewline[\doublerulesep]
\hline
$(6, 5)$  &  $1$ &\num{1.082e6}
\tabularnewline
\hline
$(7, 1)$ & $2$ & \num{-1.124e6}
\tabularnewline
$(7, 2)$ & $3$ &\num{-1.138e6}
\tabularnewline
$(7, 3)$ & $4$ & \num{1.113e6} 
\tabularnewline
$(7, 4)$ & $5$ & \num{1.115e6}  
\tabularnewline
$(7, 5)$ & $6$ & \num{1.112e6}  
\tabularnewline
$(7, 6)$ & $7$ & \num{1.106e6}  
\tabularnewline
$(7, 70)$ & $8$ & \num{-1.103e6}  
\tabularnewline
$(7, 71)$ & $9$ & \num{-1.111e6}  
\tabularnewline
$(7, 72)$ & $10$ & \num{-1.118e6}  
\tabularnewline
\hline
$(15, 43)$ & $11$ & \num{3.604e-12}  
\tabularnewline
$(15, 44)$ & $12$ & \num{2.891e-12}  
\tabularnewline
$(15, 45)$ & $13$ & \num{2.591e-12}  
\tabularnewline
$(15, 46)$ & $14$ & \num{1.693e-12}  
\tabularnewline
$(15, 48)$ & $15$ & \num{2.272e-12}  
\tabularnewline
$(15, 49)$ & $16$ & \num{4.359e-12}  
\tabularnewline
$(15, 50)$ & $17$ & \num{7.439e-12}  
\tabularnewline
$(15, 51)$ & $18$ & \num{1.032e-11}  
\tabularnewline
\hline
$(16, 44)$ & $19$ & \num{-1.258e5}  
\tabularnewline
\hline

\end{tabular}
\end{table}

A final remark pertains the computational complexity 
of the proposed time-varying
detection algorithm, which is mainly dominated by matrix
inversions in \eqref{eq:mmse} and \eqref{eq:sinr-ex},
both involving $\mathcal{O}(S^3)$ flops
to be repeated for each value of $u$ and $m$.
However, through straightforward manipulations, omitted here for brevity, it can be proven that
$\widetilde{\bm F}^{(m)}_k[u]$ can be implemented as the cascade of a time-invariant MMSE filter followed by a time-varying Doppler frequency shift compensation stage, i.e., 
$\widetilde{\bm F}_k^{(m)}[u]  = \{\bm E_k^{(m)}[u]\}^* \, {\bm F}_k$,
where 
\be
\bm F_k \eqdef \bm H_k^{\herm} \, (\bm H_k \, \bm H_k^{\herm} + \sigma_{w}^2 
\, \bm I_{S})^{-1} \in \Cset^{L_k \times S}
\ee
with
$\bm H_k \eqdef \bm P \, \bm V_k \, \bm A_k/\sqrt{\rho_k} \in \Cset^{S \times L_k}$ and the diagonal matrix 
$\bm E_k^{(m)}[u] \in \Cset^{L_k \times L_k-m}$ derived from $\bm E_k^{(m-1)}[u]$ by removing its $[j_{\text{max}}^{(m-1)}]$-th column, with $\bm E_k^{(0)}[u] = \bm E_k[u]$.
Such a decomposition
entails a reduction of the computational load
in evaluating \eqref{eq:mmse}, since
matrix inversion can be performed only when the iterative algorithm starts and
the only time-varying operation boils down to a multiplication by the diagonal
matrix $\{\bm E^{(m) }_k[u]\}^*$, which is performed at each iteration.
As regards SINR calculation in \eqref{eq:sinr-ex},
it can be proven that \eqref{eq:sinr-ex} turns out to be
time invariant and, moreover, 
matrix inversion can be
done once, i.e., at the start of the iterative algorithm.

\section{Sum-rate capacity analysis}
\label{sec:capacity}

Assuming that the satellites 
encode the information using 
an i.i.d. Gaussian code\footnote{If the channel input symbols are chosen from a discrete constellation, 
e.g., a quadrature amplitude modulation (QAM) constellation of size $Q$, the corresponding information rate $\capa^{\text{$Q$-QAM}}_{k}$
is upper-bounded by $\capa_k$. The rate reduction caused by the use of $Q$-QAM constellations 
depends on the SINR of the channel \cite{Bellorado_2006}. For high SINR values, indeed, the rate reduction may become very large since 
$\capa_k$ grows logarithmically with SINR, whereas $\capa^{\text{$Q$-QAM}}_{k}$ is strictly upper-bounded by the logarithm of the constellation 
size $Q$. For low SINR values, instead, $\capa^{\text{$Q$-QAM}}_{k}$ approaches 
$\capa_k$.}, 
recalled that \eqref{eq:sinr-ex} is time-invariant,
the sum rate of the $k$-th group $\mathcal{L}_k$
is given by 
\be
\capa_{k} \eqdef \rho_{k} 
\sum_{m=0}^{L_k-1} 
\log_2\left(1+
\text{SINR}^{(m)}_{k, \ell_{\text{max}}^{(m)}} \right)
\quad \text{(in bits/s/Hz)}
\label{eq:u-rate}
\ee
for $k \in \{1,2,\ldots, G\}$,
where $\text{SINR}^{(m)}_{k, \ell}$ has been defined in 
\eqref{eq:sinr-ex} (with $k$ omitted).
After tedious but straightforward algebraic manipulations, it can
verified that 
\barr
\capa_{k} & = \rho_{k} \log_2 \text{det} 
\left( \bm I_{S} + \frac{1}{\sigma_{w}^2} 
\widetilde{\bm{H}}_k[u] \, \widetilde{\bm{H}}_k^\herm[u]
\right) \nonumber \\ & =
\rho_{k} \log_2 \text{det} 
\left( \bm I_{S} + \frac{1}{\sigma_{w}^2 \, \rho_{k}} \,
\bm{P} \, \bm{V}_k \, \bm{A}_k \, \bm{A}_k^* \, \bm{V}_k^\herm
\, \bm{P}^\trasp  \right) \: .
\label{eq:u-rate2}
\earr
We observe that \eqref{eq:u-rate2}
corresponds to the maximum sum rate that can be achieved by the considered system.
This confirms the well-known fact that SC at the transmitters
and MMSE-SIC at the receiver allow to achieve the boundary of the MAC 
capacity region \cite{Tse-book}.
The  {\em sum-rate capacity} of the whole satellite system can be expressed as follows
\be
\capaa_\text{sum} = \sum_{k=1}^{G}  
\rho_{k} \log_2 \text{det} 
\left( \bm I_{S} + \frac{1}{\sigma_{w}^2 \, \rho_{k}} \,
\bm{P} \, \bm{V}_k \, \bm{A}_k \, \bm{A}_k^* \, \bm{V}_k^\herm
\, \bm{P}^\trasp  \right)  \: .
\label{eq:u-rate-tot}
\ee

We provide in the following two subsections a comparative performance analysis 
in terms of sum-rate capacity of 
three different multiple access schemes, i.e., pure-NOMA, pure-OMA, and
hybrid NOMA-OMA, by unveiling the crucial role played by 
the Doppler frequency shifts of the satellites that contend for channel resources.
To this aim, for the sake of simplicity, we assume that 
$\bm{P}$ is diagonally dominant with real nonzero diagonal entries
$p_0, p_1, \ldots, p_{S-1}$.

\subsection{Comparison between pure-NOMA and pure-OMA schemes}
\label{sec:comp-oma-noma}

Eq.~\eqref{eq:u-rate-tot} subsumes the sum-rate capacity of pure-NOMA 
and pure-OMA schemes as special cases.
Indeed,  the sum-rate capacity of the pure-NOMA technique  can be obtained from
\eqref{eq:u-rate-tot} by setting $G=1$, which implies $L_1=L$
and $\rho_1=1$, 
i.e., NOMA is implemented among 
all the $L$ satellites with feasible links, hence yielding
\be
\capaa_\text{sum}^\text{NOMA}   = \log_2 \text{det} 
\left( \bm I_{S} + \frac{1}{\sigma_{w}^2}  \,
\bm{P} \, \bm{V} \, \bm{A} \, \bm{A}^* \, \bm{V}^\herm
\, \bm{P}  \right)  
\label{eq:u-rate-tot-noma}
\ee
where 
$\bm{V} \eqdef  [\bm{v}_1, \bm{v}_2, \ldots, \bm{v}_L] \in\Cset^{S \times L}$ collects all the Vandermonde vectors associated to the $L$ satellites 
and  
$\bm{A} \eqdef \text{diag}[A_1, A_2, \ldots, A_L] $ gathers all the corresponding channel gains on its main diagonal.
Let $\bC \eqdef \bm{P} \, \bm{V} \, \bm{A} =[\bc_1, \bc_2, \ldots, \bc_L] \in \Cset^{S \times L}$,
eq.~\eqref{eq:u-rate-tot-noma} can be rewritten as follows
\be
\capaa_\text{sum}^\text{NOMA}    = \log_2 \text{det} 
\left( \bm I_{S} + \frac{1}{\sigma_{w}^2}  \sum_{\ell=1}^L \bc_\ell \, \bc_\ell^\herm \right) \: .
\label{eq:u-rate-tot-noma-2}
\ee
Since the matrices 
$\bm{P}$ and $\bm{A}$ are non-singular, we underline that 
$\text{rank}(\bC)=\text{rank}(\bm{V})$. 

On the other hand, the sum-rate capacity of the pure-OMA technique can be obtained from
\eqref{eq:u-rate-tot} by setting $G=L$, 
which implies $L_k=1$, i.e., the 
$L$ satellites with feasible links access the channel in an
orthogonal manner,  thus obtaining
\be
\capaa_\text{sum}^\text{OMA}  = \sum_{\ell=1}^{L}  
\rho_{\ell} \log_2 \text{det} 
\left( \bm I_{S} + \frac{1}{\sigma_{w}^2 \, \rho_{\ell}} \,
\bc_\ell \, \bc_\ell^\herm \right)  \:.
\label{eq:u-rate-tot-oma}
\ee
By virtue of the matrix determinant lemma \cite{Horn2012}, eq.~\eqref{eq:u-rate-tot-oma} 
can be equivalently rewritten as 
\barr
\capaa_\text{sum}^\text{OMA} & =
\sum_{\ell=1}^{L}  \rho_{\ell} \log_2 \left( 1+ \frac{\|\bm{c}_\ell\|^2}{\sigma_{w}^2 \, \rho_{\ell}} \right) 
\nonumber \\ & =
\sum_{\ell=1}^{L}  \rho_{\ell} \log_2 \left( 1+ \energy_\text{p} \, \frac{|A_\ell|^2}{\sigma_{w}^2 \, \rho_{\ell}}\right) 
\label{eq:u-rate-tot-oma-2}
\earr
where $\energy_\text{p} \eqdef \sum_{s=0}^{S-1} p^2_s$.
It is worth noticing that, in the pure-OMA case, the sum-rate capacity does not depend on the Doppler frequency shifts of the satellites.

Since the determinant is a log-concave function on the set of positive definite matrices, it 
follows from Jensen's inequality applied directly to \eqref{eq:u-rate-tot-oma}
that $\capaa_\text{sum}^\text{OMA} \le 
\capaa_\text{sum}^\text{NOMA}$, where we have also accounted for \eqref{eq:rhok}.
The difference between  $\capaa_\text{sum}^\text{NOMA}$ and 
$\capaa_\text{sum}^\text{OMA}$ is {\em exactly} zero if $\bC$
has rank one, that is, $\text{rank}(\bm{V})=1$,
which happens when the $L$ satellites with feasible ISLs towards the sink (see Definition~\ref{def:feasible}) have exactly the same 
Doppler frequency shifts, i.e., $\overline{\nu} \eqdef \nu_1 = \nu_2 = \cdots =\nu_L$. Indeed, in such a 
particular case, it results that $\bc_\ell = A_\ell \, \bm{P} \, \overline{\bv}$, 
$\forall \ell \in \{1,2,\ldots, L\}$, with 
$\overline{\bm v} \eqdef [1,e^{j 2 \pi \overline{\nu}}, \ldots, 
e^{j 2 \pi \overline{\nu} (S-1)}]^{\trasp} \in \Cset^{S}$,
and, thus, 
\be
\capaa_\text{sum}^\text{OMA}= 
\capaa_\text{sum}^\text{NOMA} = \log_2 \left( 1+ \frac{\energy_\text{p}}{\sigma_w^2} \sum_{\ell=1}^L 
|A_\ell|^2 \right)
\ee
provided that the DoF fractions of the pure-OMA scheme in \eqref{eq:u-rate-tot-oma}
are chosen as 
\be
\rho_\ell^\text{OMA} =\frac{|A_\ell|^2}
{\displaystyle \sum_{q=1}^L |A_q|^2} 
\label{eq:rho_opt-oma}
\ee
for $\ell \in \{1,2,\ldots, L\}$. As intuitively expected, 
in the case of equal Doppler frequency shifts, the largest sum-rate capacity is achieved 
with pure-OMA. 
However, condition $\nu_1 = \nu_2 = \cdots =\nu_L$ does not  hold 
in practice since inter-OP ISLs exhibit significantly different 
Doppler frequency shifts (see Table~\ref{tab:doppler} and
the discussion in Subsection~\ref{sec:discuss})
and, in this context,  pure-OMA is largely suboptimal in
terms of sum-rate capacity.

\subsection{Comparison between pure-NOMA and hybrid NOMA-OMA}
\label{sec:comp-noma-hybridmomaoma}

A natural question arises about the advantage of performing a partition of 
the $L$ satellites with feasible ISLs towards the sink in groups $\mathcal{L}_1, \mathcal{L}_2,
\ldots, \mathcal{L}_G$ over the pure-NOMA option.
In this respect, let $\bC_k \eqdef \bm{P} \, \bm{V}_k \, \bm{A}_k \in \Cset^{S \times L_k}$
and observe that $\bC_1, \bC_2, \ldots, \bC_G$ are matrices each gathering 
a different subset of the columns in $\bC$, which will be referred to as 
{\em column matrices} of $\bC$.

By using again the log-concavity of the determinant in the set 
of positive definite matrices, the following inequality comes 
from the application of the Jensen's inequality to \eqref{eq:u-rate-tot}:
\be
\capaa_\text{sum} \le  \log_2 \text{det} 
\left( \bm I_{S} + \frac{1}{\sigma_{w}^2}  
\sum_{k=1}^{G}   \bC_k \, \bC_k^\herm \right)  =\capaa_\text{sum}^\text{NOMA}
\label{eq:u-rate-tot-ineq}
\ee
where we have used  \eqref{eq:rhok} and noticed that, since  
$\bC_1, \bC_2, \ldots, \bC_G$ are column matrices of $\bC$, one has
$\sum_{k=1}^{G}   \bC_k \, \bC_k^\herm=\bC \, \bC^\herm$. 
Henceforth, the multiple access strategy where the $L$ satellites are divided into groups 
of $L_k$ satellites with NOMA within each group and OMA between the groups
is suboptimal in terms of sum-rate capacity. 
Equality in \eqref{eq:u-rate-tot-ineq} holds when the matrix  $\bC_k \, \bC_k^\herm$
differs only for a scalar real constant, i.e., $\bC_k \, \bC_k^\herm 
= \chi_k \, {\bm \Omega}$, $\forall k \in \{1,2,\ldots, G\}$, which in its turn
would be fulfilled in the scenario where 
satellites belonging to the same group $\mathcal{L}_k$ experience the same 
path losses, i.e., 
$|A_{\ell_{k,1}}|^2=|A_{\ell_{k,2}}|^2=\cdots= |A_{\ell_{k,L_k}}|^2=\chi_k$, 
$\forall k \in \{1,2,\ldots, G\}$, and, moreover, the groups
$\mathcal{L}_1, \mathcal{L}_2,
\ldots, \mathcal{L}_G$ gather satellites with the same Doppler frequency shifts,
i.e., $\bm{V}_k=\bm{V}_k'$, for each $k \neq k'$. One has
$\capaa_\text{sum}=\capaa_\text{sum}^\text{NOMA}$, provided that 
the DoF fractions of the hybrid NOMA-OMA scheme in \eqref{eq:u-rate-tot} are 
chosen as $\rho_k = \chi_k/\sum_{q=1}^{G} \chi_q$. 

The performance gain of the pure-NOMA scheme over the 
hybrid NOMA-OMA one comes at the price of computation complexity, 
which in a pure-NOMA scheme grows with the number $L$ of satellites
having feasible ISLs with respect to the sink. Moreover, the available DoF are limited by the 
dimension $S$ of the observable space \cite{Tse-book} and, thus, 
there is no further DoF gain beyond
having $S$ satellites performing pure-NOMA concurrently.
Interestingly, there is another aspect to be accounted for the specific scenario at hand. 
We have seen from Table~\ref{tab:doppler} that some satellites with feasible ISLs 
towards the sink have the same Doppler frequency shift. 
Therefore, {\em what happens if there exist 
satellites with the same Doppler frequency shift that access the channel by
using the pure-NOMA scheme?}
According to \eqref{eq:u-rate-tot-ineq}, the pure-NOMA technique
ensures the best possible sum-rate capacity even in this particular case.
However, {\em what are the individual rates of the satellites with
the same Doppler frequency shifts?}
To give an answer,  let us consider the two-satellite example, 
where the received signal \eqref{eq:rbar} at the sink ends up to 
\be
\overline{\bm{r}}[u]=\widetilde{\bm{h}}_{1}[u] \, b_1[u]
+\widetilde{\bm{h}}_{2}[u] \, b_2[u] + \overline{\bm{w}}[u]
\ee 
with $\widetilde{\bm{h}}_{\ell}[u] \eqdef A_{\ell} 
\, e^{j 2 \pi S \nu_{\ell} u}
\, \bm{P} \, \bm{v}_{\ell} \in \Cset^S$, for $\ell \in \{1,2\}$
(the dependence on the group index $k$ has been
omitted since $G=1$ in the pure-NOMA scheme).  
Assuming without loss of generality that satellite $1$
is canceled first,  the individual rates of the two satellites  are given by 
$\capa_\ell=\log_2(1+ \text{SINR}_\ell)$ (in bits/s/Hz), for $\ell \in \{1,2\}$, where,
starting from \eqref{eq:sinr-ex}, after some algebraic manipulations, one has
\barr
\text{SINR}_1 & \eqdef \frac{ \sigma_w^2 \, |A_1|^2 \, \energy_\text{p} +
|A_1|^2 \, |A_2|^2 \, \left(\energy_\text{p}^2 - \left|\bm{v}_{1}^\herm \, 
\bm{P}^2 \, \bm{v}_{2}\right|^2\right)}{\sigma_w^2 \, 
\left(|A_2|^2 \, \energy_\text{p}+\sigma_w^2\right)}
\label{eq:SINR_1}
\\
\text{SINR}_2 & \eqdef |A_2|^2 \, \frac{\energy_\text{p}}{\sigma_w^2}
\label{eq:SINR_2}
\earr
with
\be
\left|\bm{v}_{1}^\herm \, 
\bm{P}^2 \, \bm{v}_{2}\right|^2 = \left| \sum_{s=0}^{S-1} p^2_s  \, e^{-j \, 2 \pi s (\nu_1-\nu_2)}\right|^2 \le \energy_\text{p}^2 \: .
\label{eq:ineq-mod2}
\ee
Equality in \eqref{eq:ineq-mod2} holds when the two satellites have the same
Doppler frequency shift, i.e., $\nu_1=\nu_2$. Hence, we can infer
\be
\text{SINR}_1 \ge \text{SINR}_1^\text{min} \eqdef  \frac{|A_1|^2}
{\displaystyle |A_2|^2
+\frac{\sigma_w^2}{\energy_\text{p}}}
\label{eq:SINR-ineq}
\ee
where $\text{SINR}_1^\text{min}$ is the SINR of satellite $1$ in the case 
when the Doppler frequency shifts are equal. Two interesting conclusions can be drawn from 
\eqref{eq:SINR-ineq}. First, if the two satellites have different Doppler frequency shifts,
i.e., $\nu_1 \neq \nu_2$,  the satellite that is decoded first can benefit of a 
larger rate, compared to the case when $\nu_1=\nu_2$. Second, in the
high signal-to-noise ratio (SNR) regime, i.e., as 
$\energy_\text{p}/\sigma_w^2 \to + \infty$, the rate 
of satellite $2$ increases without bound, whereas 
the rate of satellite $1$ tends 
approximately to 
$1$ bit/s/Hz when, besides having the same Doppler frequency shift,  
the satellites experience comparable path losses, i.e., 
$|A_1| \approx |A_2|$.

In a nutshell, even though the pure-NOMA scheme exhibits the highest sum-rate capacity,
it might result in a very unfair resource allocation, since  
basically the satellites 
having different Doppler frequency shifts might be allowed to achieve 
higher data rates with respect to 
those exhibiting equal Doppler frequency shifts. 
This suggests that, compared to the pure-NOMA scheme,  
the hybrid NOMA-OMA strategy
may better capture fairness among satellites by judiciously 
partitioning them in groups on the basis of their 
Doppler frequency shifts, 
without appreciably degrading the sum-rate capacity. 
Next section studies such an insight 
in greater detail.

\section{Optimization of the hybrid NOMA-OMA scheme}
\label{sec:opt}

Along the same lines of Subsection~\ref{sec:discuss}, 
we observe that $r_k \eqdef \text{rank}(\bC_k)=\text{rank}(\bm{V}_k) \le \min(S, L_k)$,
where the equality holds if condition \eqref{eq:cond4} is fulfilled, 
i.e., the Doppler frequency shifts of the satellites belonging to 
$\mathcal{L}_k$ are distinct.
Let $\mu_{k,1} \ge \mu_{k,2} \ge \cdots \ge \mu_{k,r_k} >0$
be the nonzero singular values of $\bm{C}_k$, eq.~\eqref{eq:u-rate-tot} can be written as 
\be
\capaa_\text{sum} = \sum_{k=1}^{G}  \sum_{s=1}^{r_k}
\rho_{k} \, \log_2 
\left( 1+ \frac{\mu_{k,s}^2}{\sigma_{w}^2 \, \rho_{k}}\right)  
\label{eq:u-rate-tot-2}
\ee
which shows that all the available per-group DoF
can be exploited when $r_k$ reaches its maximum value, given by $\min(S, L_k)$,
i.e., the Doppler frequency shifts of the satellites belonging to 
$\mathcal{L}_k$ are distinct, for each $k \in \{1,2,\ldots, G\}$.
Consequently, we impose by design that 
the partition $\mathbb{P}(\mathcal{L})$ obeys $r_k=\min(S, L_k)$.

There are two design issues of 
the hybrid NOMA-OMA scheme to be faced with:
(i) distribution of the available DoF among the satellites
having feasible ISLs with respect to the sink, i.e., optimization of the variables
$\rho_1, \rho_2, \ldots, \rho_G$, subject to constraint \eqref{eq:rhok};
(ii) choice of the partition $\mathbb{P}(\mathcal{L})$, i.e., 
setting of $G$ and definition of 
the satellite groups
$\mathcal{L}_1, \mathcal{L}_2, \ldots, \mathcal{L}_G$, 
subject to constraint $r_k=\min(S, L_k)$.
Such optimization objectives are coupled to each other as shown soon after.

As a first step, by using the product rule of logarithms, 
we equivalently rewrite \eqref{eq:u-rate-tot-2} as follows
\be
\capaa_\text{sum} = \sum_{k=1}^{G}  
\rho_{k} \, \log_2 
\left( \prod_{s=1}^{r_k}
\left[1+ \frac{\mu_{k,s}^2}{\sigma_{w}^2 \, \rho_{k}}\right] \right)  \: .
\label{eq:u-rate-tot-3}
\ee
As a second step, by invoking the arithmetic-geometric mean inequality for non-negative real numbers 
\cite{Horn2012}, one obtains the following upper bound:
\be
\capaa_\text{sum} \le \sum_{k=1}^{G}  
r_k \, \rho_{k} \, \log_2 
\left( 1+  
\frac{1}{\sigma_{w}^2 \, \rho_{k}} 
\left[ \frac{1}{r_k} \sum_{s=1}^{r_k} \mu_{k,s}^2 \right]
\right)
\label{eq:u-rate-tot-ub}
\ee
with equality if and only if the nonzero singular values of 
$\bC_k$ are all equal, i.e., 
the (spectral) condition number $\kappa(\bC_k) 
\eqdef \mu_{k,1}/\mu_{k,r_k}$ is equal to $1$ \cite{Horn2012},
for each $k \in \{1,2,\ldots, G\}$.
The upper bound in \eqref{eq:u-rate-tot-ub} cannot be achieved
in practice since 
$\kappa(\bC_k)=1$ if and 
only if 
$\bC_k \,  \bC_k^\herm \propto \bm I_{S}$ and $S \le L_k$ or 
$\bC_k^\herm \, \bC_k\propto \bm I_{L_k}$
and $S \ge L_k$. 
Henceforth, for the problem at hand, $\kappa(\bC_k)>1$
and, to maximize the sum-rate capacity,  
the design objective of the satellite groups is
to ensure that 
$\bC_1,\bC_2, \ldots, \bC_G $
have condition numbers as close to $1$ as possible.
Moreover, recalling that 
the available DoF are limited by $S$, we design the 
partition $\mathbb{P}(\mathcal{L})$ such that each group 
has a number of satellites $L_k$ smaller than or equal to $S$, i.e., 
we impose that  $r_k =\min(S, L_k)=L_k$, for each $k \in \{1,2,\ldots, G\}$.
To summarize, we came up with 
the following guideline for the design of the satellite groups
$\mathcal{L}_1, \mathcal{L}_2, \ldots, \mathcal{L}_G$.

\begin{mdframed}
{\em Guideline $\text{D}_{\text{g}}$ (Design of satellite groups)}:
$ \forall k \in \{1,2,\ldots, G\}$, form the group $\mathcal{L}_k$
by selecting from the set $\mathcal{L}$, which gathers all the satellites
with feasible ISLs towards the sink, $L_k$ satellites such that:

\begin{itemize}

\item 
their Doppler frequency shifts are distinct, i.e., $r_k=\text{rank}(\bC_k)=L_k$;

\item
$\kappa(\bC_k)$ is close to $1$ as possible.

\end{itemize}

\end{mdframed}

The reformulation of the guideline $\text{D}_{\text{g}}$ in terms of an 
algorithm with manageable complexity is pursued in Subsection~\ref{sec:anticlust}.
For the time being, let us assume that $\text{D}_{\text{g}}$ has been fulfilled such
that we can regard the upper bound in \eqref{eq:u-rate-tot-3} as a sharp 
approximation of the sum-rate capacity, i.e., 
\be
\capaa_\text{sum} \approx \sum_{k=1}^{G}  
L_k \, \rho_{k} \, \log_2 
\left( 1+  
\frac{1}{\sigma_{w}^2 \, L_k \, \rho_{k}} 
\sum_{s=1}^{L_k} \mu_{k,s}^2  
\right) \:.
\label{eq:u-rate-tot-approx}
\ee
Given the group sizes $L_1, L_2, \ldots, L_G$, with 
$ \sum_{k=1}^{G}  L_k =L$, the problem of finding the values of  
the DoF fractions $\rho_1, \rho_2, \ldots, \rho_G$, 
for which the approximation \eqref{eq:u-rate-tot-approx} of 
$\capaa_\text{sum}$ is maximized, subject to  \eqref{eq:rhok}, 
does not admit  a closed-form solution. 
For such a reason, as a third step, we resort to a max-min approach. 
Specifically, we propose to maximize 
with respect to $\rho_1, \rho_2, \ldots, \rho_G$,
subject to  \eqref{eq:rhok}, 
the lower bound 
\be
\capaa_\text{sum} \gtrapprox L_\text{min} \sum_{k=1}^{G}  
 \rho_{k} \, \log_2 
\left( 1+  
\frac{1}{\sigma_{w}^2 \, L_k \, \rho_{k}} 
\sum_{s=1}^{L_k} \mu_{k,s}^2  
\right) \:.
\label{eq:u-rate-tot-approx-2}
\ee
with $L_\text{min} \eqdef \min_{k \in \{1,2,\ldots,G\}} L_k$,
where the approximation holds when the satellite groups have the
same cardinality, i.e., $L_k$ is independent of $k$.
By Jensen's inequality \cite{Horn2012}, we have
\begin{multline}
\sum_{k=1}^{G}  
 \rho_{k} \, \log_2 
\left( 1+  
\frac{1}{\sigma_{w}^2 \, L_k \, \rho_{k}} 
\sum_{s=1}^{L_k} \mu_{k,s}^2  
\right)  \\ \le 
\log_2 
\left(1+\frac{1}{\sigma_{w}^2 } 
\sum_{k=1}^{G}  \frac{1}{L_k} \sum_{s=1}^{L_k}  \mu_{k,s}^2 
\right)
\end{multline}
where the upper bound is achieved when 
the DoF fractions 
are chosen as follows 
\be
\rho_k^\text{hybrid} = \frac{\displaystyle \frac{1}{L_k} \sum_{s=1}^{L_k} \mu_{k,s}^2 }
{\displaystyle \sum_{q=1}^{G}  \frac{1}{L_q}  \sum_{s=1}^{L_q} \mu_{q,s}^2 } 
\label{eq:rho_opt-hybrid}
\ee
for $k \in \{1,2,\ldots, G\}$, which represent the solution of the proposed max-min problem. 
Having dealt with design issue (i), we provide in the forthcoming subsection a solution of issue (ii) according to $\text{D}_{\text{g}}$.

\subsection{Choice of the satellite groups}
\label{sec:anticlust}

Preliminarily, we observe that, since $\bm P$ 
and $\bm{A}_k$ are diagonal non-singular matrices, 
and $\bm{V}_k$ is full rank under guideline $\text{D}_{\text{g}}$,
one has
\barr
1 < \kappa(\bm{C}_k) & \le \kappa(\bm{P}) \, 
\kappa( \bm{V}_k) \, \kappa(\bm{A}_k)
\nonumber \\ & = 
\frac{p_\text{max}}{p_\text{min}} \, 
\frac{A_{k,\text{max}}}{A_{k,\text{min}}} \, 
\kappa( \bm{V}_k)
\label{eq:ineq-cond-1}
\earr
with
$p_\text{min} \eqdef \min_{s \in \{0, 1, \ldots, S-1\}} |p_s|$,
$p_\text{max}  \eqdef \max_{s \in \{0, 1, \ldots, S-1\}} |p_s|$,
$A_{k,\text{min}} \eqdef \min_{q \in \{1, 2, \ldots, L_k\}} |A_{\ell_{k,q}}|$, 
$A_{k,\text{max}}  \eqdef \max_{q \in \{1, 2, \ldots, L_k\}} |A_{\ell_{k,q}}|$.
We note that $p_\text{min}$ and $p_\text{max}$ depend
on the choice of the pulse-shaping filter $p(t)$ and sampling period $T_\text{c}$,
and are not functions of $k$.
On the other hand, $A_{k,\text{min}}$ and $A_{k,\text{max}}$
depend on the geometric path losses of
the satellites belonging to the group $\mathcal{L}_k$, 
and they do not depend on the corresponding 
phase misalignments.
The condition number of the Vandermonde matrix $\bm{V}_k$ depends
on the Doppler frequency shifts of the satellites belonging to $\mathcal{L}_k$.\footnote{If
the Doppler frequency shifts in $\mathcal{L}_k$
lay on a fixed grid with width $1/S$ when $S \ge L_k$, then 
one would have a perfectly conditioned Vandermonde matrix, i.e., 
$\kappa( \bm{V}_k)=1$.}
Since the singular values of a matrix are continuously
depending on its entries, we expect that $\kappa( \bm{V}_k)$ 
grows if such Doppler frequency shifts become close to each other.
To characterize such a behavior, a meaningful notion of distance between 
the Doppler frequency shifts is necessary. Since $\bm{V}_k$ is a Vandermonde matrix 
on the unit circle, we can resort to 
the (normalized) arc length between points on the unit circle.

\begin{definition}[{\em Wrap-around distance}]
\label{def:distance}
{\em For $k \in \{1,2,\ldots, G\}$, let us define the {\em Doppler frequency shift set}
$\mathcal{D}_k \eqdef \{\nu_{\ell_{k,1}}, \nu_{\ell_{k,2}}, \ldots, \nu_{\ell_{k,L_k}}\}$
of the $k$-th satellite group $\mathcal{L}_k$. The wrap-around distance between two Doppler frequency shifts
 $\nu_{\ell_{k,q}}, \nu_{\ell_{k,q'}}\in \mathcal{D}_k$ is defined as follows
\be
|\nu_{\ell_{k,q}}-\nu_{\ell_{k,q'}}|_{\mathcal{D}_k} \eqdef 
\min_{\eta \in \mathbb{Z}} |\nu_{\ell_{k,q}}-\nu_{\ell_{k,q'}}+\eta| \: .
\ee
The {\em minimal separation distance} $\delta_k$ of the Doppler frequency shift set  
$\mathcal{D}_k $ is given by        
\be
\delta_k \eqdef \min_{
\shortstack{\footnotesize $\nu_{\ell_{k,q}}, \nu_{\ell_{k,q'}}\in \mathcal{D}_k$ 
\\ \footnotesize $\nu_{\ell_{k,q}} \neq \nu_{\ell_{k,q'}}$}} 
|\nu_{\ell_{k,q}}-\nu_{\ell_{k,q'}}|_{\mathcal{D}_k} \: .
\ee
}
\end{definition}

It should be noted that, under assumption 
$f_{k,\text{max}} \, T \ll 1$ (see Subsection~\ref{sec:rec-signal}), 
it follows that $\delta_k \ll {1}/{S}$, i.e., the Doppler frequency shifts  
in $\mathcal{D}_k$  are {\em nearly colliding}.
Inequality \eqref{eq:ineq-cond-1} and Definition~\ref{def:distance}
suggest the following reformulation of $\text{D}_{\text{g}}$. 

\begin{mdframed}
{\em Guideline $\text{D}_{\text{g}}$ (Reformulation)}:
$ \forall k \in \{1,2,\ldots, G\}$, form the group $\mathcal{L}_k$
by selecting from the set $\mathcal{L}$, which gathers all the satellites
with feasible ISLs towards the sink, $L_k$ satellites such that:

\begin{itemize}

\item 
their Doppler frequency shifts are distinct, i.e., $r_k=\text{rank}(\bC_k)=L_k$;

\item 
the minimal separation distance $\delta_k$ is maximized;

\item
their path losses are comparable, i.e., $A_{k,\text{max}} \approx A_{k,\text{min}}$.

\end{itemize}

\end{mdframed}

On the basis of the discussion carried out in 
Subsection~\ref{sec:comp-noma-hybridmomaoma}, the first two requirements
(i.e., $\text{rank}(\bC_k)=L_k$ and maximization of $\delta_k$) avoid
an unfair resource allocation within the group $\mathcal{L}_k$. 
Since the individual rates of the satellites also depend on their
corresponding path losses, the third requirement $A_{k,\text{max}} \approx A_{k,\text{min}}$
goes on the direction of improving fairness among satellites belonging to the same group, too.

A strictly exhaustive procedure for finding the partition $\mathbb{P}(\mathcal{L})$
according to $\text{D}_{\text{g}}$ is often computationally unmanageable, even for small
values of $L$. 
In the subsequent two subsections, we propose two algorithms:
in the former one, referred to as {\em Doppler-based partitioning},
we propose to partition the satellites into groups with the aim of creating 
high Doppler heterogeneity within each group  and high Doppler similarity between groups
({\em anticlustering}),
thus accounting for the first two requirements of $\text{D}_{\text{g}}$,
without however considering the third requirement in terms of path losses; 
in the latter one, referred to as {\em max-fairness partitioning},
we propose to maximize a global fairness index, which implicitly 
accounts  for all the three requirements of $\text{D}_{\text{g}}$,
through an exhaustive search in suitably-reduced search space.

\begin{algorithm}
\caption{Doppler-based partitioning and subsequent DoF fraction assignment.}
\label{alg:alg_1}
\begin{algorithmic}
\vspace{3mm}
\STATE  {\bf Input quantities}: The set $\mathcal{L}$ of the
$L$ satellites with feasible ISLs towards the sink within the considered observation
interval and their corresponding Doppler frequency shifts 
$f_1, f_2, \ldots, f_L$.

\STATE
{\bf Output quantities}: The partition $\mathbb{P}({\mathcal{L}}) = \{\mathcal{L}_k\}_{k=1}^{G}$ 
of $\mathcal{L}$.

\begin{enumerate}[1.]

\itemsep=1mm

\item
Set $G$ equal to the number of satellites belonging \\ 
to the same OP of the sink (they have equal Doppler frequency shifts
very close to zero) and 
arbitrarily assign satellites to the $G$ anticlusters by ensuring that each anticluster 
consists of (approximately) the same number of satellites.

\item
Select the first satellite 
and check how \eqref{eq:Var-max}
will \\ change if the selected item is swapped with each satellite that is currently
assigned to a different anticluster.

\item
After performing each possible swap, realize
the one that entails  
the most significant increase of \eqref{eq:Var-max}.
No swap is realized if the objective cannot be improved.

\item
The procedure terminates when the swap operation 
\\ has
been repeated for each satellite.

\item
Assign the DoF fractions to each group according to 
\eqref{eq:rho_opt-hybrid} ({\em optimized DoF assignment}) or 
set $\rho_k^\text{hybrid}=1/G$ ({\em uniform DoF assignment}).

\end{enumerate}
\end{algorithmic}
\end{algorithm}

\subsubsection{Doppler-based partitioning}
\label{sec:anti-cluster}

Anticlustering consists of maximizing 
instead of minimizing a clustering objective function \cite{Val1983,Val1998}. 
For Doppler-based partitioning, we aim at maximizing 
the variance among Doppler features of the satellites with feasible ISLs
towards the sink (so-called {\em $k$-means anticlustering}).
Specifically, we propose to determine the 
partition $\mathbb{P}(\mathcal{L})$ by maximizing
the cost function:
\begin{equation}
\mathcal{V}(\mathcal{L}_1, \mathcal{L}_2, \ldots, \mathcal{L}_G) \eqdef
\sum_{k=1}^G \sum_{q=1}^{L_k}
\left(f_{\ell_{k,q}}-\overline{f}_k \right)^2
\label{eq:Var-max}
\end{equation}
where $\overline{f}_k  \eqdef
\frac{1}{L_k}\sum_{q=1}^{L_k} f_{\ell_{k,q}}$ is the {\em cluster centroid}
of the $k$-th group.
It is shown in \cite{Spa1986} that 
there is a direct connection between 
$\mathcal{V}(\mathcal{L}_1, \mathcal{L}_2, \ldots, \mathcal{L}_G)$
and the location of the cluster centroids.
If the cost function \eqref{eq:Var-max} is maximal, the cluster centroids 
are as close as possible to the overall centroid
\be
\overline{f}  \eqdef
\frac{1}{L} \sum_{k=1}^G \sum_{q=1}^{L_k} f_{\ell_{k,q}}
\ee
and, therefore, to each other \cite{Spa1986}. Thus, $k$-means anticlustering directly optimizes
the similarity of the mean attribute values between clusters.
Unfortunately, finding a partitioning that maximizes Doppler heterogeneity
according to this criterion is still computationally challenging (i.e., it can be
shown to be NP-hard). Indeed, the number of anticlustering partitions 
increases exponentially with $L$, quickly rendering it impossible to find  
them all out in acceptable running time. For such a reason, we propose a 
low-complexity solution, which is based on a swap-based heuristic \cite{Pap2021}.

Our heuristic algorithm is based on swapping satellites between anticlusters such 
that each swap improves the objective function \eqref{eq:Var-max}
by the largest possible margin.
The proposed algorithm is summarized as Algorithm~\ref{alg:alg_1}
at the top of next page. It can be shown \cite{Pap2021} 
that Algorithm~\ref{alg:alg_1} performs very similarly 
to exact solution methods,
e.g.,
integer linear programming, which ensure
global optimal solution at the price of 
an exponential explosion of the running time
even for small values of $L$.
However, Algorithm~\ref{alg:alg_1} does not explicitly 
account for system fairness and the partition 
is determined without accounting for the available DoF.

\begin{figure*}
\centering
\includegraphics[width=\textwidth]{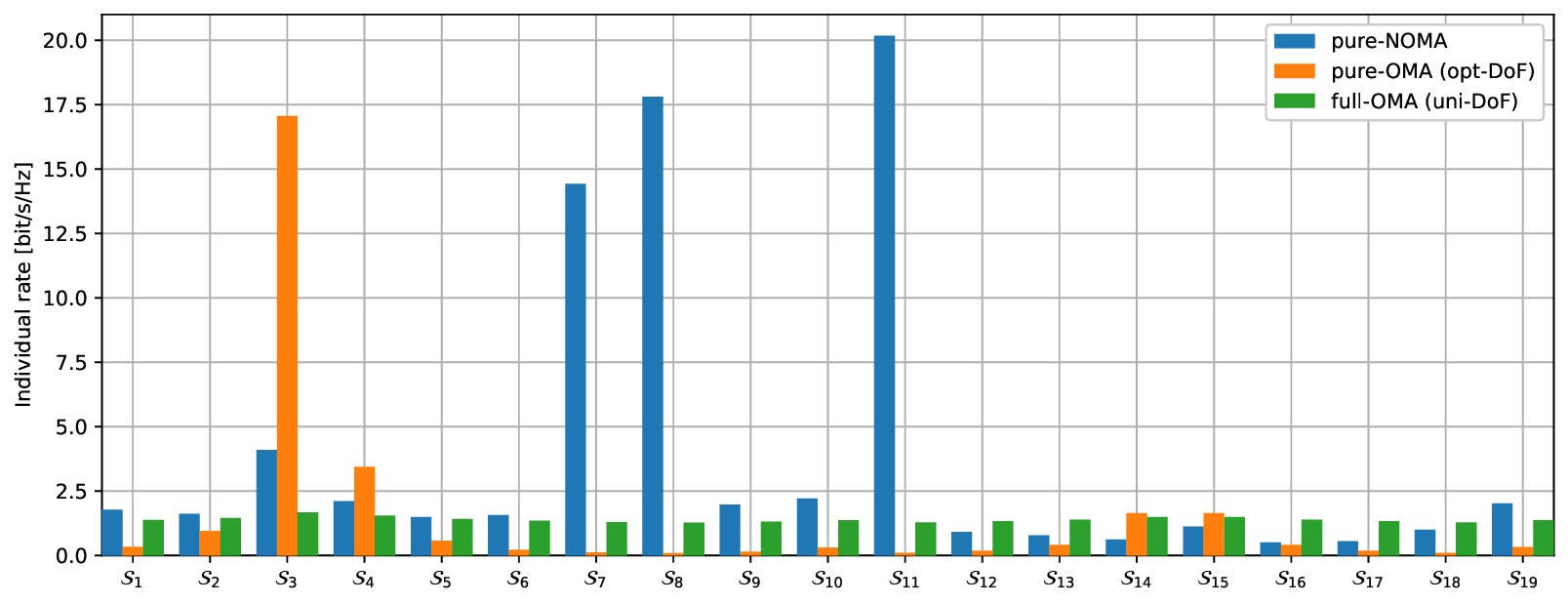}
\caption{Individual rates of the $L=19$ satellites with feasible ISLs towards the sink of pure-NOMA and pure-OMA schemes.}
\label{fig:num-res-noma-oma}
\end{figure*}

\subsubsection{Max-fairness partitioning}
\label{sec:max-fair}

A simple yet informative index to quantify 
the fairness of a scheduling scheme was proposed in \cite{Jain.1998},
which, according to \eqref{eq:u-rate}, is given by \eqref{eq:fair-index}.
\begin{figure*}[!t]
\normalsize
\be
\fair(\mathcal{L}_1, \mathcal{L}_2, \ldots, \mathcal{L}_G)  \eqdef \frac{\displaystyle \left[\sum_{k=1}^G 
\sum_{m=0}^{L_k-1} \rho_k \log_2\left(1+
\text{SINR}^{(m)}_{k, \ell_{\text{max}}^{(m)}} \right)\right]^2}
{\displaystyle L \sum_{k=1}^G 
\sum_{m=0}^{L_k-1} \left[ \rho_k  \log_2\left(1+
\text{SINR}^{(m)}_{k, \ell_{\text{max}}^{(m)}}\right)\right]^2} \: .
\label{eq:fair-index}
\ee
\hrulefill
\end{figure*}
If all the $L$ satellites get the same rate, then the fairness index is $1$
and the multiple access method is said to be $100\%$ fair. On the other hand,
a multiple access scheme that favors only a few selected users
has a fairness index tending to $0$. 

We propose to maximize \eqref{eq:fair-index} by resorting to a \emph{reduced} exhaustive search.
The possible partitions of the set $\mathcal{L}$ is the Bell number 
$B_{L} = \frac{1}{e} \sum_{i=0}^{+\infty} \frac{i^L}{i!}$.\footnote{
It can be interpreted as the $L$-th moment of a Poisson distribution with expected value 1.}
For the example of Fig.~\ref{fig:fig_4}, the size of the set 
$\mathcal{L}$ is $L=19$ and, in this case, there are 
$B_{19}=5.832.742.205.057$ partitions to be generated and tested.  
To reduce the set of candidate partitions to a manageable size, 
the proposed procedure is based on a problem-specific preprocessing step that 
we call {\em prepartitioning}.
In the prepartitioning step, we preliminarily identify the satellites that exhibit Doppler frequency
shifts that are as similar as possible, by exploiting constellation network information.
Specifically, according to Table~\ref{tab:doppler} (see also Subsection~\ref{sec:discuss}),
we know that the satellites with feasible ISLs belonging to the same OP of the sink
are characterized by Doppler frequency shifts almost equal to zero. 
Such satellites are assigned to different $G$ groups and, thus, they operate according to
the OMA scheme. So doing, the problem boils down to assign the remaining $L-G$
satellites to the $G$ groups such that to maximize \eqref{eq:fair-index}, whose 
number of possible partitions is given by $G^{L-G}$. In the case of 
Fig.~\ref{fig:fig_2} (see also Table~\ref{tab:doppler}), one has $G=8$
and, hence, there are $8^{19-8}=8.589.934.592$ candidate partitions,
which can be generated and tested in a reasonable running time.
The proposed algorithm is summarized as Algorithm~\ref{alg:alg_2}
at the top of this page.
It is noteworthy that Algorithm~\ref{alg:alg_2} jointly performs
partitioning and DoF assignment to the satellite groups.

\vspace{2mm}
{\em Remark}:
Algorithms~\ref{alg:alg_1} and \ref{alg:alg_2}
can be implemented in a full LEO 
context-aware and reconfigurable architecture \cite{Karapetyan2015,Yan2021, Roth2021}, 
where each satellite
is capable of collecting information from the neighboring ones.
Specifically, by relying on the knowledge 
of the relative positions of the satellites as well as on the characteristics
of the receiver equipment, each satellite can build a \emph{feasibility map} 
- which can be calculated offline - containing the time windows over which 
it can establish a feasible communication link with other neighboring satellites.
Standard Request-to-Send (RTS) and Clear-to-Send (CTS) mechanisms 
can be employed to setup the communication links between 
the relevant satellites and the sink. 

\section{Numerical results}
\label{sec:num}

In this section, we provide numerical results aimed at
corroborating the developed performance analysis 
and comparing the performance of the three
multiple access schemes, namely, pure-OMA, 
pure-NOMA, and hybrid NOMA-OMA. 
To this aim, with reference to an observation interval of length $0.5$ ms, 
we consider the same 
simulation setting used to generate 
Fig.~\ref{fig:fig_4} and Table~\ref{tab:doppler}
(see  Subsections~\ref{sec:example} and \ref{sec:discuss} for details).
We recall that, in this scenario, the number of satellites with 
feasible ISLs towards the sink is $L=19$, which, according 
to Figs.~\ref{fig:fig_2} and \ref{fig:fig_3}, represents the 
more challenging application from the MAC viewpoint.\footnote{Results
not reported here for the lack of space show that similar conclusions
can be drawn for other values 
of $L$ reported in Fig.~\ref{fig:fig_2}. 
}

Unless 
otherwise specified, the receiving device at the sink is affected by
thermal noise with a noise figure $F$ of 8 dB
and the oversampling factor at the sink is $S=8$.
As performance metrics, we consider 
both the sum-rate capacity and the fairness index \eqref{eq:fair-index}.

With reference to the pure-OMA scheme, we report the performance
when DoF fractions are chosen according to \eqref{eq:rho_opt-oma},
referred to as ``pure-OMA (opt-DoF)", 
as well as when they are allocated uniformly, i.e. $\rho_1=\rho_2=\ldots=\rho_L=1/L$,
referred to as ``pure-OMA (uni-DoF)".
In a similar way, we implement the two versions of the proposed 
hybrid NOMA-OMA technique optimized as in 
Algorithms~\ref{alg:alg_1} and \ref{alg:alg_2} by assigning 
the DoF fractions in accordance with both 
\eqref{eq:rho_opt-hybrid}, referred to as 
``hyb-NOMA-OMA-1 (opt-DoF)"
and ``hyb-NOMA-OMA-2 (opt-DoF)", respectively, 
and the uniform rule $\rho_1=\rho_2=\ldots=\rho_G=1/G$, 
referred to as 
``hyb-NOMA-OMA-1 (uni-DoF)"
and ``hyb-NOMA-OMA-2 (uni-DoF)", respectively.

\begin{algorithm}
\caption{Max-fairness joint partitioning and DoF fraction assignment.}
\label{alg:alg_2}
\begin{algorithmic}
\vspace{3mm}
\STATE  {\bf Input quantities}: The set $\mathcal{L}$ of the
$L$ satellites with feasible ISLs towards the sink within the considered observation
interval and their corresponding Doppler frequency shifts 
$f_1, f_2, \ldots, f_L$.

\STATE
{\bf Output quantities}: The partition $\mathbb{P}({\mathcal{L}}) = \{\mathcal{L}_k\}_{k=1}^{G}$ 
of $\mathcal{L}$.

\begin{enumerate}[1.]

\itemsep=1mm

\item
Individuate the $G$ satellites $\{\ell_{1,1}, \ell_{2,1}, \ldots, \ell_{G,1}\} \subset
\mathcal{L}$ that belong 
to the same OP of the sink (they have equal Doppler frequency shifts
very close to zero) and perform the prepartitioning 
$\mathcal{L}_1=\{\ell_{1,1}\}, 
\mathcal{L}_2=\{\ell_{2,1}\}, \ldots, \mathcal{L}_G=\{\ell_{G,1}\}$.

\item
Generate a candidate solution by assigning the remaining $L-G$
satellites to the $G$ group.

\item
Assign the DoF fractions to each candidate group according to 
\eqref{eq:rho_opt-hybrid} ({\em optimized DoF assignment}) or 
set $\rho_k^\text{hybrid}=1/G$ ({\em uniform DoF assignment}).

\item
Evaluate the fairness index \eqref{eq:fair-index}.

\item
After performing Steps~2, 3, and 4 for all the possible partitions, choose
the one that maximizes \eqref{eq:fair-index}.

\end{enumerate}
\end{algorithmic}
\end{algorithm}

\begin{table}
\centering
\scriptsize
\caption{Sum-rate capacity and fairness index of
all of the analyzed schemes for $F = 8$ dB.}
\label{tab:num-res-noma-oma_8}
\begin{tabular}{ccc}
\hline
\noalign{\vskip\doublerulesep}
\textbf{Algorithm} & $\boldsymbol{\capaa}_\text{\textbf{sum}}$ &
$\boldsymbol{\fair}$
\tabularnewline[\doublerulesep]
\hline
pure-NOMA & \num{76.645} & \num{0.316}
\tabularnewline
pure-OMA (opt-DoF) & \num{28.237} & \num{0.136}
\tabularnewline
pure-OMA (uni-DoF) & \num{26.409} & \num{0.995}
\tabularnewline
Alg.~1 (opt-DoF) & \num{61.697} & \num{0.384}
\tabularnewline
Alg.~2 (opt-DoF) & \num{58.065} & \num{0.800}
\tabularnewline
Alg.~1 (uni-DoF) & \num{53.100} & \num{0.954}
\tabularnewline
Alg.~2 (uni-DoF) & \num{54.856} & \num{0.997}
\tabularnewline
\hline
\end{tabular}
\end{table}

\begin{table*}[htbp]
\centering
\scriptsize
\caption{Partition of the $L=19$ satellites with feasible ISLs towards the sink
of the hybrid NOMA-OMA scheme.}
\label{tab:num-res-part}
\begin{tabular}{cc}
\hline
\noalign{\vskip\doublerulesep}
\textbf{Scheme} & \textbf{Partition} 
 \tabularnewline[\doublerulesep]
\hline
\textbf{Alg.~1 (opt-DoF/uni-DoF)} & 
$\mathcal{L}_1=\{1,9\}, \mathcal{L}_2=\{2,4,11\}, \mathcal{L}_3=\{3,6,12\},
\mathcal{L}_4=\{5,10,13\}, \mathcal{L}_5=\{7,17\}, \mathcal{L}_6=\{8,14\},
\mathcal{L}_7=\{15,19\}, \mathcal{L}_8=\{16,18\}$
\tabularnewline
\textbf{Alg.~2 (opt-DoF)} &  
$\mathcal{L}_1=\{1,3,6,7,8,11,19\}, \mathcal{L}_2=\{2,12\}, \mathcal{L}_3=\{13\},
\mathcal{L}_4=\{5,14\}, \mathcal{L}_5=\{10,15\}, \mathcal{L}_6=\{16\},
\mathcal{L}_7=\{17\}, \mathcal{L}_8=\{4,9,18\}$
\tabularnewline
\textbf{Alg.~2 (uni-DoF)} &
$\mathcal{L}_1=\{6,11\}, \mathcal{L}_2=\{1,12\}, \mathcal{L}_3=\{9,13\},
\mathcal{L}_4=\{3,14,19\}, \mathcal{L}_5=\{4,8,15\}, \mathcal{L}_6=\{7,16\},
\mathcal{L}_7=\{10,17\}, \mathcal{L}_8=\{2,5,18\}$   
\tabularnewline
\hline
\end{tabular}
\end{table*}

\begin{figure*}
\centering
\includegraphics[width=\textwidth]{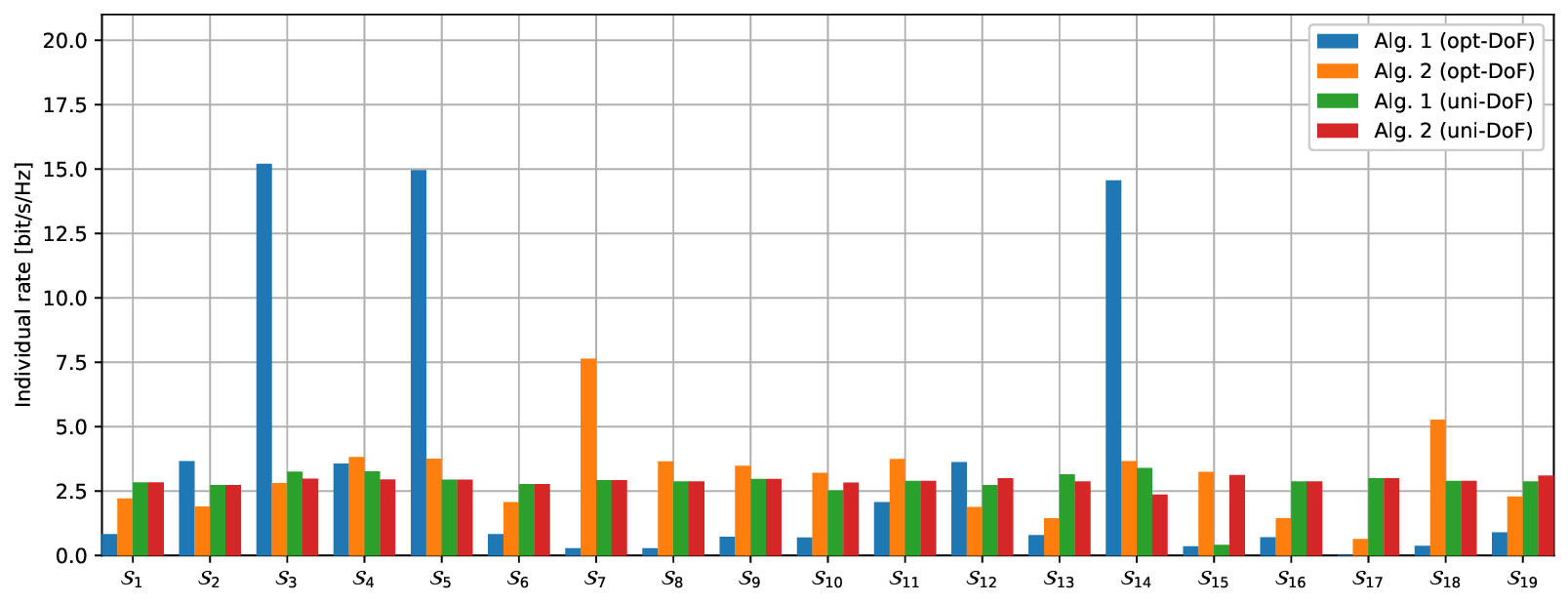}
\caption{Individual rates of the $L=19$ satellites with feasible ISLs towards the sink of the hybrid NOMA-OMA schemes.}
\label{fig:num-res-hybrid}
\end{figure*}

Fig.~\ref{fig:num-res-noma-oma} reports the individual rates of 
all the satellites $\{\mathcal{S_\ell}\}_{\ell=1}^L$
with feasible ISLs towards the sink
of the pure-NOMA and pure-OMA schemes. On the other hand, 
the individual rates of 
all the satellites with feasible ISLs towards the sink
of the hybrid NOMA-OMA schemes are shown in Fig.~\ref{fig:num-res-hybrid}.
The corresponding synthetic parameters, i.e.,
the sum-rate capacity and the fairness index,
of all of the considered multiple access schemes are
summarized in Table~\ref{tab:num-res-noma-oma_8}.
It is seen that,
besides significantly outperforming the pure-OMA technique
in terms of sum-rate capacity
(see Subsection~\ref{sec:comp-oma-noma}), 
the pure-NOMA scheme also ensures a better fairness among the 
satellites when the pure-OMA DoF fractions are chosen as in 
\eqref{eq:rho_opt-oma}.
This is in accordance with \cite{Shin2017}.
If the pure-OMA  DoF fractions are allocated uniformly, the 
sum-rate capacity of the pure-OMA scheme further cuts down
but, however, it becomes almost $100 \%$ fair.
Corroborating the analysis of 
Subsection~\ref{sec:comp-noma-hybridmomaoma}, 
the results of Table~\ref{tab:num-res-noma-oma_8}
confirm the fact that, even though it is 
optimal in an information-theoretical sense, the pure-NOMA
schemes is quite unfair when the transmissions of satellites 
with similar Doppler frequency shifts are superimposed.

Table~\ref{tab:num-res-noma-oma_8} also highlights 
that Alg.~1 (opt-DoF), which performs
a Doppler-based partitioning by using \eqref{eq:rho_opt-hybrid},
substantially ensures the same fairness index of the pure-NOMA
scheme. However, when the hybrid NOMA-OMA scheme
is designed by using Algorithm~\ref{alg:alg_1} with uniform 
DoF fractions, Alg.~1 (uni-DoF) is as fair as pure-OMA (uni-DoF), 
by remarkably assuring a sum-rate capacity increase of about 
$27$ bits/s/Hz.
Algorithm~\ref{alg:alg_2} allows  the hybrid NOMA-OMA scheme
to enhance network fairness at the price of a slight reduction
of the sum-rate capacity compared to
Alg.~1 (opt-DoF) and Alg.~1 (uni-DoF). For instance, 
Alg.~1 (opt-DoF)  enhances the fairness index of  
$50 \%$ with respect to pure-NOMA scheme, while 
outperforming pure-OMA (opt-DoF) of about $30$ bits/s/Hz,
which is the best trade-off between fairness and information rate
for the methods under comparison.

Table~\ref{tab:num-res-part} gathers the partitions found by the different 
algorithms. It should be remembered that the partition provided 
by Algorithm~\ref{alg:alg_1} does not depend on the DoF fractions
and, thus, Alg.~1 (opt-DoF)  and Alg.~1 (uni-DoF) return the same partitions.
It can be observed that the number of groups is $G=8$, which
is exactly the number of satellites with feasible ISLs belonging to the
same OP of the sink (see Table~\ref{tab:doppler}). Moreover, it is noteworthy that the proposed algorithms 
ensure a reasonable small number of satellites for each group, which allows
to compensate error propagation effects of practical implementations of
the SIC process by slightly increasing the coding block length 
(see also Subsection~\ref{sec:discuss}). 

\begin{figure}
	\centerline{\includegraphics[width=0.8\columnwidth]{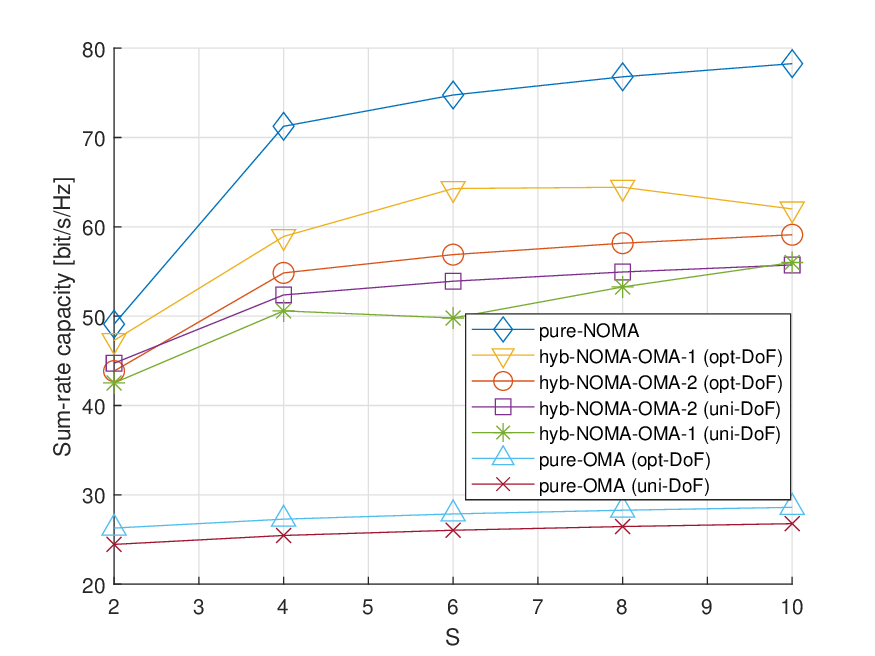}}
	\caption{Sum-rate capacity of the considered schemes as a function of $S$.}
	\label{fig:fig_5}
\end{figure}

\begin{figure}
	\centerline{\includegraphics[width=0.8\columnwidth]{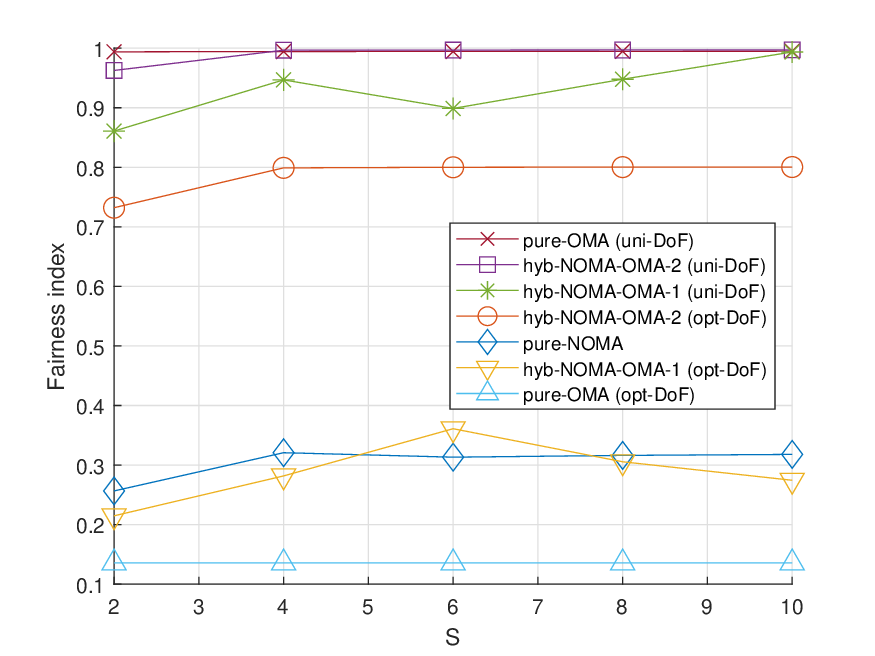}}
	\caption{Fairness index of the considered schemes as a function of $S$.}
	\label{fig:fig_6}
\end{figure}

Figs.~\ref{fig:fig_5} and \ref{fig:fig_6} depict the sum-rate capacity 
and the fairness index of the schemes under comparison as a function 
of the oversampling factor $S$ at the sink.
It can be argued that $S$ has a large impact on the sum-rate capacity of
the NOMA-based schemes, whereas the rate performance 
of the OMA-based techniques 
slightly improves for increasing values of $S$. In particular, 
the sum-rate capacity of hyb-NOMA-OMA-1 (opt-DoF)
tends to diminish when the oversampling rate at the sink
becomes greater than the number $G=8$ of the groups.  
On the other hand, 
the fairness index is slight influenced by the values of $S$
for all the considered schemes, especially for the 
pure-NOMA and pure-OMA techniques. 

Finally, Tables~\ref{tab:num-res-noma-oma_16} and \ref{tab:num-res-noma-oma_4}
report the sum-rate capacity and the fairness index of 
the considered multiple access schemes for the two different values 
of the noise figure $F=16$ dB and $F=4$ dB, respectively. 
Compared to the results of Table~\ref{tab:num-res-noma-oma_8}, it can
be inferred that the noise figure has a negligible impact on
the fairness index.
On the other hand, as expected, the sum-rate capacity is
very sensitive to changes of the noise figure. However, 
the above conclusions drawn from the comparative analysis
carried out with $F=8$ still hold for such different values 
of the noise figure.

\section{Conclusions}
\label{sec:conc}

In this paper, to ensure
high spectral efficiency and massive connectivity in LEO constellations, 
we have studied the feasibility of 
superimposing the transmissions of 
satellites with different Doppler frequency profiles
in a NOMA setting.
To enhance network fairness, we have developed a theoretical framework that
allows one to partition a pool of satellites into groups using orthogonal channel
resources, with the goal of achieving high within-group heterogeneity in the 
Doppler frequency domain such that concurrent transmissions of satellites
belonging to the same group can be
separated at the sink through MMSE-SIC reception.  
Our numerical results show that the proposed hybrid NOMA-OMA
scheme significantly outperforms existing OMA ones
in terms of sum-rate capacity, by exhibiting a better trade-off
between rate and fairness than the capacity-achieving pure-NOMA technique.
Such a trade-off becomes significant in massive LEO systems, where
a fair share of the scarce radio spectrum is one of the main concern.


\appendix[The ECEF coordinate system]
\label{sec:app}

\begin{figure}
    \centering
    \includegraphics[width=\columnwidth]{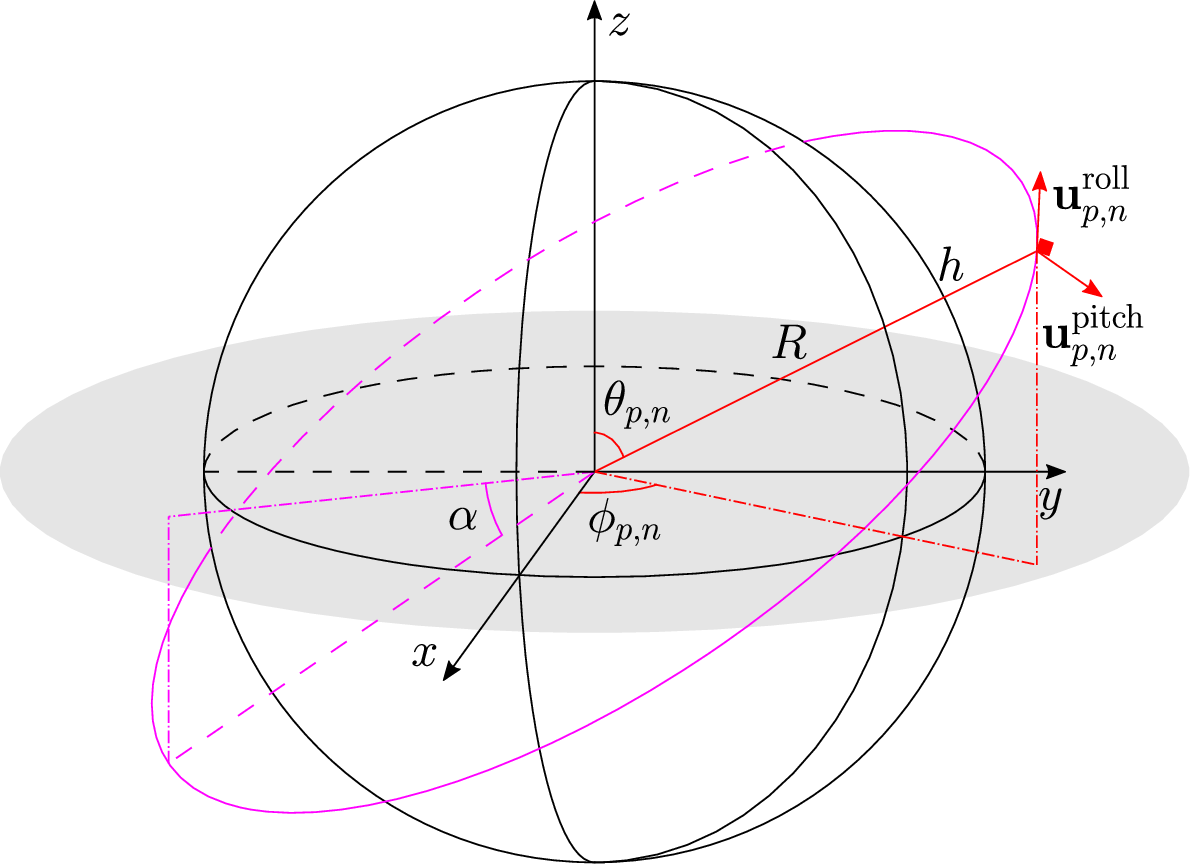}
    \caption{The ECEF coordinate system. Both
    Cartesian and spherical coordinates are highlighted in the picture.}
    \label{fig:ecef}
 \end{figure}

With reference to Fig.~\ref{fig:ecef}, 
the position in the ECEF coordinate system 
of the $n$-th satellite
belonging to the $p$-th orbit plane,
with $p \in \{1,2,\ldots,P\}$ and
$n \in \{1,2,\ldots,N\}$,
is defined by the following rotations:
\begin{equation}
\mathbf{r}_{p,n}(t) \eqdef
\begin{bmatrix}
x_{p,n}(t) \\ y_{p,n}(t) \\ z_{p,n}(t)
\end{bmatrix} =
(R+h) \, \mathbf{\Xi}_p \, 
\mathbf{\Pi} \, \mathbf{e}_{p,n}(t)
\end{equation}
where $\mathbf{e}_{p,n}(t) \eqdef
[\cos(\omega t + \gamma_{p,n}),
\sin(\omega t + \gamma_{p,n}), 0]^\trasp$,
\begin{equation}
\mathbf{\Pi} \eqdef
\begin{bmatrix}
\cos\alpha & 0 & -\sin\alpha \\
0 & 1 & 0 \\
\sin\alpha & 0 & \cos\alpha
\end{bmatrix}
\end{equation}
and 
\begin{equation}
\mathbf{\Xi}_p \eqdef
\begin{bmatrix}
\cos\beta_p & -\sin\beta_p & 0 \\
\sin\beta_p & \cos\beta_p & 0 \\
0 & 0 & 1
\end{bmatrix}
\end{equation}
with 
\barr
\beta_p & \eqdef \frac{2 \,\pi \, (p-1)}{P}
\\
\gamma_{p,n} & \eqdef \frac{2 \,\pi\, (n-1)}{N} + \frac{2 \,\pi \,F \,(p-1)}{K}
\earr
$\omega$ and $\alpha \in [0,\pi/2]$ representing
the satellite angular speed and the orbit
inclination angle, respectively, and, 
finally, $F$ is the so-called
\emph{phasing} parameter usually designed
in order to maximize the minimum distance
between satellites \cite{Liang2021}.

The position can also be represented
in spherical coordinates, i.e.,
\begin{equation}
\mathbf{r}_{p,n}(t) =
\begin{bmatrix}
(R+h) \sin[\theta_{p,n}(t)] \cos[\varphi_{p,n}(t)] \\
(R+h) \sin[\theta_{p,n}(t)] \sin[\varphi_{p,n}(t)] \\
(R+h) \cos[\theta_{p,n}(t)]
\end{bmatrix}.
\end{equation}

\begin{table}
\centering
\scriptsize
\caption{Sum-rate capacity and fairness index of
all of the analyzed schemes for $F = 16$ dB.}
\label{tab:num-res-noma-oma_16}
\begin{tabular}{ccc}
\hline
\noalign{\vskip\doublerulesep}
\textbf{Algorithm} & $\boldsymbol{\capaa}_\text{\textbf{sum}}$ &
$\boldsymbol{\fair}$
\tabularnewline[\doublerulesep]
\hline
pure-NOMA & \num{76.645} & \num{0.316}
\tabularnewline
pure-OMA (opt-DoF) & \num{28.237} & \num{0.136}
\tabularnewline
pure-OMA (uni-DoF) & \num{23.540} & \num{0.994}
\tabularnewline
Alg.~1 (opt-DoF) & \num{68.506} & \num{0.250}
\tabularnewline
Alg.~2 (opt-DoF) & \num{58.065} & \num{0.800}
\tabularnewline
Alg.~1 (uni-DoF) & \num{46.479} & \num{0.946}
\tabularnewline
Alg.~2 (uni-DoF) & \num{48.040} & \num{0.996}
\tabularnewline
\hline
\end{tabular}
\end{table}
\begin{table}
\centering
\scriptsize
\caption{Sum-rate capacity and fairness index of
all of the analyzed schemes for $F = 4$ dB.}
\label{tab:num-res-noma-oma_4}
\begin{tabular}{ccc}
\hline
\noalign{\vskip\doublerulesep}
\textbf{Algorithm} & $\boldsymbol{\capaa}_\text{\textbf{sum}}$ &
$\boldsymbol{\fair}$
\tabularnewline[\doublerulesep]
\hline
pure-NOMA & \num{84.016} & \num{0.321}
\tabularnewline
pure-OMA (opt-DoF) & \num{30.049} & \num{0.136}
\tabularnewline
pure-OMA (uni-DoF) & \num{28.222} & \num{0.995}
\tabularnewline
Alg.~1 (opt-DoF) & \num{73.891} & \num{0.251}
\tabularnewline
Alg.~2 (opt-DoF) & \num{63.036} & \num{0.800}
\tabularnewline
Alg.~1 (uni-DoF) & \num{56.703} & \num{0.954}
\tabularnewline
Alg.~2 (uni-DoF) & \num{59.160} & \num{0.997}
\tabularnewline
\hline
\end{tabular}
\end{table}

%
%

\bibliographystyle{IEEEtran}
\bibliography{IEEEabrv,LEO_sat_bib_v2}

\begin{IEEEbiography}[
{\includegraphics[width=1in,height=1.25in,clip,keepaspectratio]{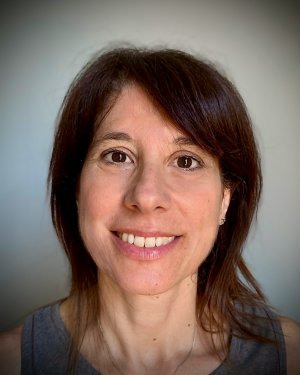}}]
{Donatella Darsena} (Senior Member, IEEE) received the Dr. Eng. degree summa cum laude in telecommunications engineering in 2001, and the Ph.D. degree in electronic and telecommunications engineering in 2005, both from the University of Napoli Federico II, Italy. From 2001 to 2002, she worked as embedded system designer in the Telecommunications, Peripherals and Automotive Group, STMicroelectronics, Milano, Italy. 
In 2005 she joined the Department of Engineering at Parthenope University of Napoli, Italy and worked first as an Assistant Professor and then as an Associate Professor from 2005 to 2022.
She is currently an Associate Professor in the Department of Electrical Engineering and Information Technology of the University of Napoli Federico II, Italy.
Her research interests are in the broad area of signal processing for communications, with current emphasis on reflected-power communications, orthogonal and nonorthogonal multiple access techniques, wireless system optimization, and physical-layer security.
Dr. Darsena has served as a Senior Editor for IEEE ACCESS since 2024, Executive Editor for IEEE COMMUNICATIONS LETTERS since 2023, and Associate Editor for IEEE SIGNAL PROCESSING LETTERS since 2020. She was an Associate Editor of IEEE ACCESS (from 2018 to 2023), of IEEE COMMUNICATIONS LETTERS (from 2016 to 2019), and Senior Area Editor of IEEE COMMUNICATIONS LETTERS (from 2020 to 2023). 
\end{IEEEbiography}

\begin{IEEEbiography}
[{\includegraphics[width=1in,height=1.25in,clip,keepaspectratio]{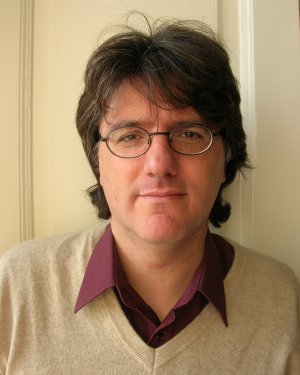}}]
{Giacinto Gelli} (Senior Member, IEEE) received the Dr. Eng. degree \textit{summa cum laude} in electronic engineering in 1990, and the Ph.D. degree in computer science and
electronic engineering in 1994, both from the University of Napoli
Federico II.

From 1994 to 1998, he was an Assistant Professor with the
Department of Information Engineering, Second University of
Napoli.
Since 1998 he has been with the Department of Electrical Engineering and Information Technology, University of Napoli Federico II,
first as an Associate Professor,
and since November 2006 as a Full Professor of telecommunications.
He also held teaching positions at the University of Napoli Parthenope.
His research interests are in the broad area of
signal and array processing for communications,
with current emphasis on reflected-power communication systems, 
multicarrier modulation systems, and
space-time techniques for cooperative and cognitive
communications systems.
\end{IEEEbiography}

\begin{IEEEbiography}
[{\includegraphics[width=1in,height=1.25in,clip,keepaspectratio]{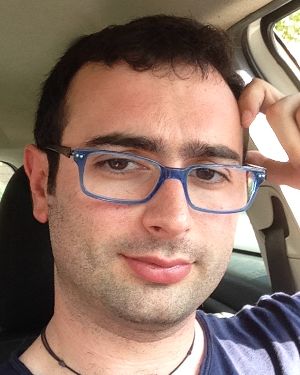}}]
{Ivan Iudice} received the B.S. and M.S. degrees
in telecommunications engineering in 2008 and 2010,
respectively, and the Ph.D. degree
in information technology and electrical engineering in 2017,
all from University of Napoli Federico II, Italy.

Since November 2011,
he has been with the Italian Aerospace Research Centre (CIRA), Capua, Italy.
He first served as part of the Electronics and Communications Laboratory
and he is currently part of the Security Unit.
He is involved in several international projects.
He serves as reviewer for several international journals
and as TPC member for several international conferences.
He is author of several papers on refereed journals and international conferences.
His research activities mainly lie in the area of
signal and array processing for communications,
with current interests focused on physical-layer security,
space-time techniques for cooperative communications systems
and reconfigurable metasurfaces.
\end{IEEEbiography}

\begin{IEEEbiography}[
{\includegraphics[width=1in,height=1.25in,clip,keepaspectratio]{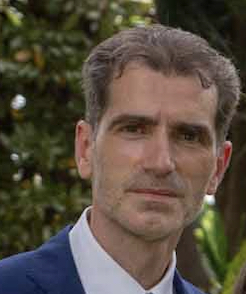}}]
{Francesco Verde}(Senior Member, IEEE)  received the Dr. Eng. degree
\textit{summa cum laude} in electronic engineering
from the Second University of Napoli, Italy, in 1998, and the Ph.D.
degree in information engineering
from the University of Napoli Federico II, in 2002.
Since December 2002, he has been with the University of Napoli Federico II, Italy. He first served as an Assistant Professor of signal theory and mobile communications
and, since December 2011, he has served as an Associate Professor of telecommunications with the Department of Electrical Engineering and Information Technology.
His research activities include reflected-power communications, 
orthogonal/non-orthogonal multiple-access techniques, wireless systems optimization, and 
physical-layer security.

Prof. Verde has been involved in several technical program committees of major IEEE conferences in signal processing and wireless communications.
He has served as Associate Editor for IEEE TRAN\-SACTIONS ON VEHICULAR TECHNOLOGY since 2022.
He was an Associate Editor of the IEEE TRANSACTIONS ON SIGNAL PROCESSING (from 2010 to 2014), IEEE SIGNAL PROCESSING LETTERS (from 2014 to 2018),
IEEE TRANSACTIONS ON COMMUNICATIONS (from 2017 to 2022), and 
Senior Area Editor of the IEEE SIGNAL PROCESSING LETTERS (from 2018 to 2023), 
as well as Guest Editor of the EURASIP Journal on Advances in Signal Processing in 2010 and SENSORS MDPI in 2018-2022.
\end{IEEEbiography}

\end{document}